%% file: sample-sigconf.tex
\documentclass[sigconf,backref=page]{acmart}
\AtBeginDocument{%
  \providecommand\BibTeX{{%
    \normalfont B\kern-0.5em{\scshape i\kern-0.25em b}\kern-0.8em\TeX}}}

\setcopyright{acmcopyright}
\copyrightyear{2024}
\acmYear{2024}
\acmDOI{XXXXXXX.XXXXXXX}

\acmConference[WSDM 24]{The 17th ACM Inernational Conference on Web Search and Data Mining}{March 04--08,
  2024}{Mérida, México}
\acmPrice{15.00}
\acmISBN{978-1-4503-XXXX-X/18/06}

\acmSubmissionID{50}

\usepackage{caption}
\usepackage{subcaption}
\usepackage{cleveref}
\usepackage{array}
\usepackage{amsthm,amsmath}

\usepackage{bbm}
\usepackage{multirow}
\usepackage{mathtools}
\usepackage[inline]{enumitem}
\usepackage{tablefootnote}
\input{math_commands.tex}

\def\vb{{\underline{ \bm{b} }}}
\def\vf{{\underline{ \bm{f} }}}
\def\vh{\underline{\bm{h}}}
\def\vp{{ \underline{\bm{p}} }}
\def\vq{{ \underline{\bm{q}} }}

\begin{document}

\title{To Copy, or not to Copy; That is a Critical Issue of the Output Softmax Layer in Neural Sequential Recommenders}

\author{Haw-Shiuan Chang}
\authornote{The work is done while the authors were at UMass.}
\email{chawshiu@amazon.com}
\author{Nikhil Agarwal}
\authornotemark[1]
\email{agnikh@amazon.com}
\affiliation{%
  \institution{Amazon.com}
  \country{USA}
}

\author{Andrew McCallum}
\affiliation{%
  \institution{University of Massachusetts}
  \streetaddress{140 Governors Dr.}
  \city{Amherst}
  \state{MA}
  \country{USA}}
\email{mccallum@cs.umass.edu}
\renewcommand{\shortauthors}{Chang, et al.}

\begin{abstract}

%repeat
Recent studies suggest that the existing neural models have difficulty handling repeated items in sequential recommendation tasks. However, our understanding of this difficulty is still limited. In this study, we substantially advance this field by identifying a major source of the problem: the single hidden state embedding and static item embeddings in the output softmax layer. Specifically, the similarity structure of the global item embeddings in the softmax layer sometimes forces the single hidden state embedding to be close to new items when copying is a better choice, while sometimes forcing the hidden state to be close to the items from the input inappropriately. To alleviate the problem, we adapt the recently-proposed softmax alternatives such as softmax-CPR to sequential recommendation tasks and demonstrate that the new softmax architectures unleash the capability of the neural encoder on learning when to copy and when to exclude the items from the input sequence. By only making some simple modifications on the output softmax layer for SASRec and GRU4Rec, softmax-CPR achieves consistent improvement in 12 datasets. With almost the same model size, our best method not only improves the average NDCG@10 of GRU4Rec in 5 datasets with duplicated items by 10\% (4\%-17\% individually) but also improves 7 datasets without duplicated items by 24\% (8\%-39\%)!

%small changes in the softmax layer could lead to large improvement;

%one of the major sources 
%a large portion of the problem comes from the output softmax layer. 
%softmax is one of the main problem

%test several alternative solutions

%sequential recommender
%Our understanding of the reasons is limited. 
%Our proposed neural sequential recommenders consistently improve SAS and GRU4Rec in 12 datasets. 
%Without significantly increasing the model size, 

%RepeatNet?

%recommend

\end{abstract}

\begin{CCSXML}




\maketitle

\input{content/introduction}

\input{content/problems}

\input{content/methods}

\input{content/experiments}

\input{content/related_work}

\section{Conclusion}
%explore and exploit
In the last decade, various neural sequential recommenders are proposed and the output softmax layer or single hidden state is used in almost all of them. These studies often focus on developing a new neural encoder architecture for some specific applications and show its superior performance on a few datasets. In our study, we show that the choice of output softmax layer is also very important in all the 12 datasets we tried. Under our experimental setup, it is even more important than the choice of the neural encoder.

In the 5 datasets without any duplication, the similar performances of \textbf{softmax + C}, \textbf{RepeatNet}, and \textbf{softmax w/o duplication~\citep{li2023repetition}} reveal that breaking the softmax bottleneck is the main source of the significant improvements of RepeatNet or removing the duplication in the post-processing in these datasets. 

Finally, we recommend setting softmax-CPR as the default method for computing the next item probability in sequential recommendation tasks due to its simplicity and universal improvement on the datasets with or without duplications.

%In our experiments, 

%are very simple and universally improve the datasets with or without duplications. Therefore, 

%suggest that  
%our analyses  

%focus mostly on developing a better neural encoder and 
%do not know why these recommenders sometimes have 
%without noticing the potential issues 

%Lots of  and focus on , which
%Most of the studies focus on 
%Lots of effort focus on improving the neural encoder. 

%The choose of softmax alternatives has a bigger effect than the choose of neural encoder.

%\todo{ The improvement is especially large when the hidden state size is small.}

%The results show that RepeatNet can improve the performance of datasets without any item duplications, and the context partition and reranker partition in Softmax-CPR further improves the performance. 

%explore and exploit.

%limited by the 
%We would like to identify where the 
%In \citet{chang2023revisiting}, the softmax-CPR overhead is much smaller compared to our time experiments. 
%We haven't optimized our Pytorch code for speed. write cuda code 

%Test in the industrial setting
%reranker step
%nearest neighbor search
%larger datasets

%The embeddings of the items in the recommendation list need to be close to the single embedding, so they are often close/similar to each other. It is interesting to see if using softmax-CPR could also increase the diversity of recommended items.

%implementation limit our speed. write cuda code
%Softmax often leads to similar .

\section{Ethical Considerations}
Modeling the repetition behavior better might sometimes intensify the filter bubble~\citep{spohr2017fake} on a recommendation-based web platform. For example, if one user keeps watching a set of videos talking about conspiracy theories, predicting the next item to be a video from this set might increase the system's accuracy, but further strengthen the intellectual isolation and polarization of the society. We believe that how to break the bubble is still an open problem and out of the scope of this paper.

%one user keeps watching conspiracy
%repetition
%keep recommending the same video, author
%bubble 
%potential negative societal impact
%Examples of negative societal impact include fairness considerations, privacy considerations, security considerations, safety considerations, misuse of the technology by malicious actors, as well as possible harms that could arise even when the technology is being used as intended and functioning correctly. 

%\section{Appendices}

\begin{acks}
We thank Tian Wang, Zihang Dai, and the anonymous reviewers for their valuable suggestions.
This work was supported 
in part by the Center for Data Science and the Center for Intelligent Information Retrieval, 
in part by the Chan Zuckerberg Initiative under the project Scientific Knowledge Base Construction, 
in part by the IBM Research AI through the AI Horizons Network, 
in part using high performance computing equipment obtained under a grant from the Collaborative R\&D Fund managed by the Massachusetts Technology Collaborative, 
and in part by the National Science Foundation (NSF) grant numbers IIS-1922090 and IIS-1763618.
Any opinions, findings, conclusions, or recommendations expressed in this material are those of the authors and do not necessarily reflect those of the sponsor.
\end{acks}

\bibliographystyle{ACM-Reference-Format}
\bibliography{sample-base}

\appendix
\input{content/appendix}

\end{document}

%% file: math_commands.tex
\usepackage{amsmath,amsfonts,bm}

\def\eqref#1{equation~\ref{#1}}
\def\1{\bm{1}}

\DeclareMathAlphabet{\mathsfit}{\encodingdefault}{\sfdefault}{m}{sl}
\SetMathAlphabet{\mathsfit}{bold}{\encodingdefault}{\sfdefault}{bx}{n}

%% file: content/introduction.tex
\section{Introduction}

\begin{figure}[t!]
\centering
\includegraphics[width=0.9\linewidth]{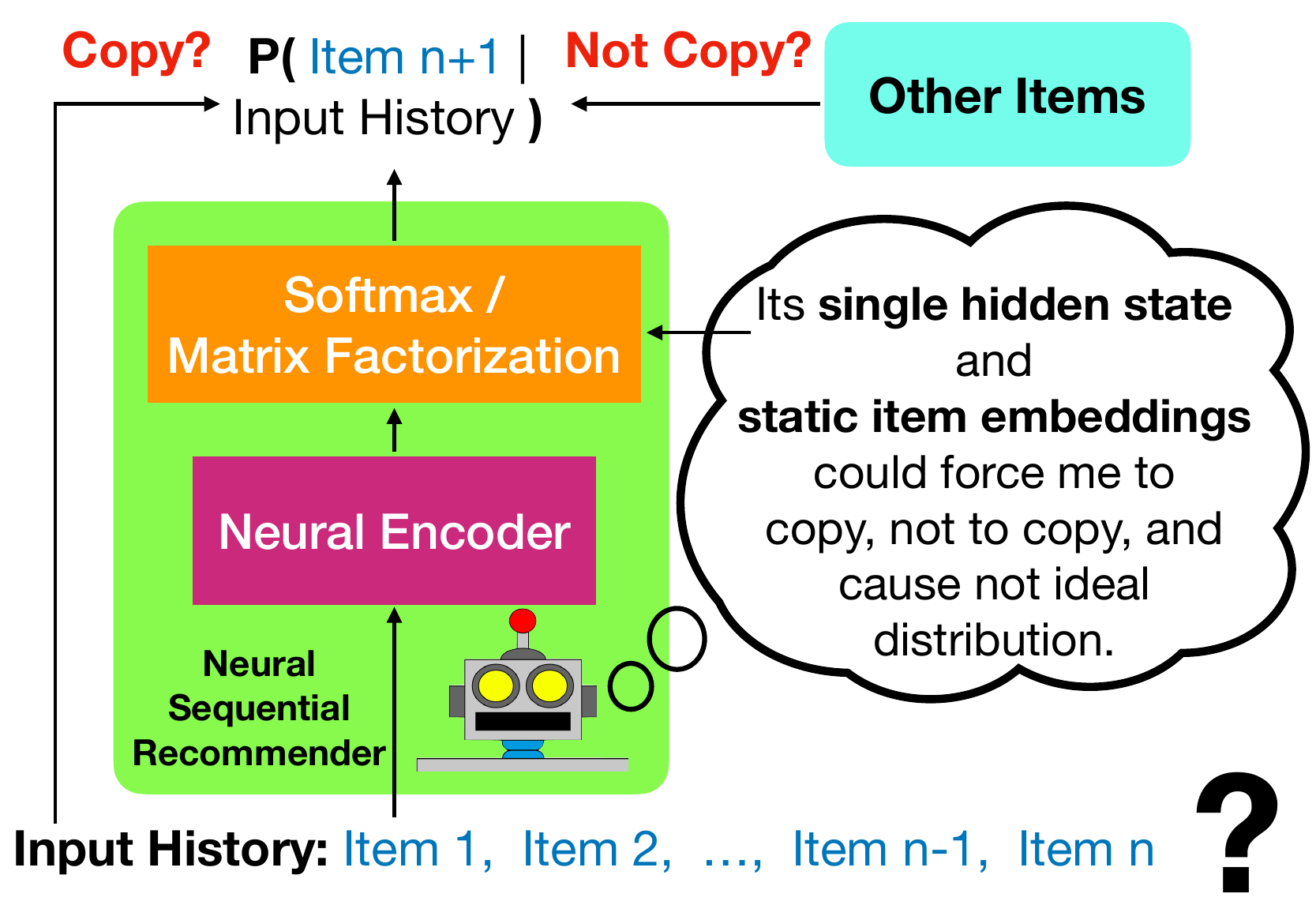}
\caption{The output softmax layer which is used in most neural networks prevents the recommender from modeling the ideal next item distribution and the softmax bottleneck also makes the recommender unable to learn the correct copying behavior from the training data. }
\label{fig:first_page}
\end{figure}

\begin{figure*}[t!]
\centering
\includegraphics[width=0.9\linewidth]{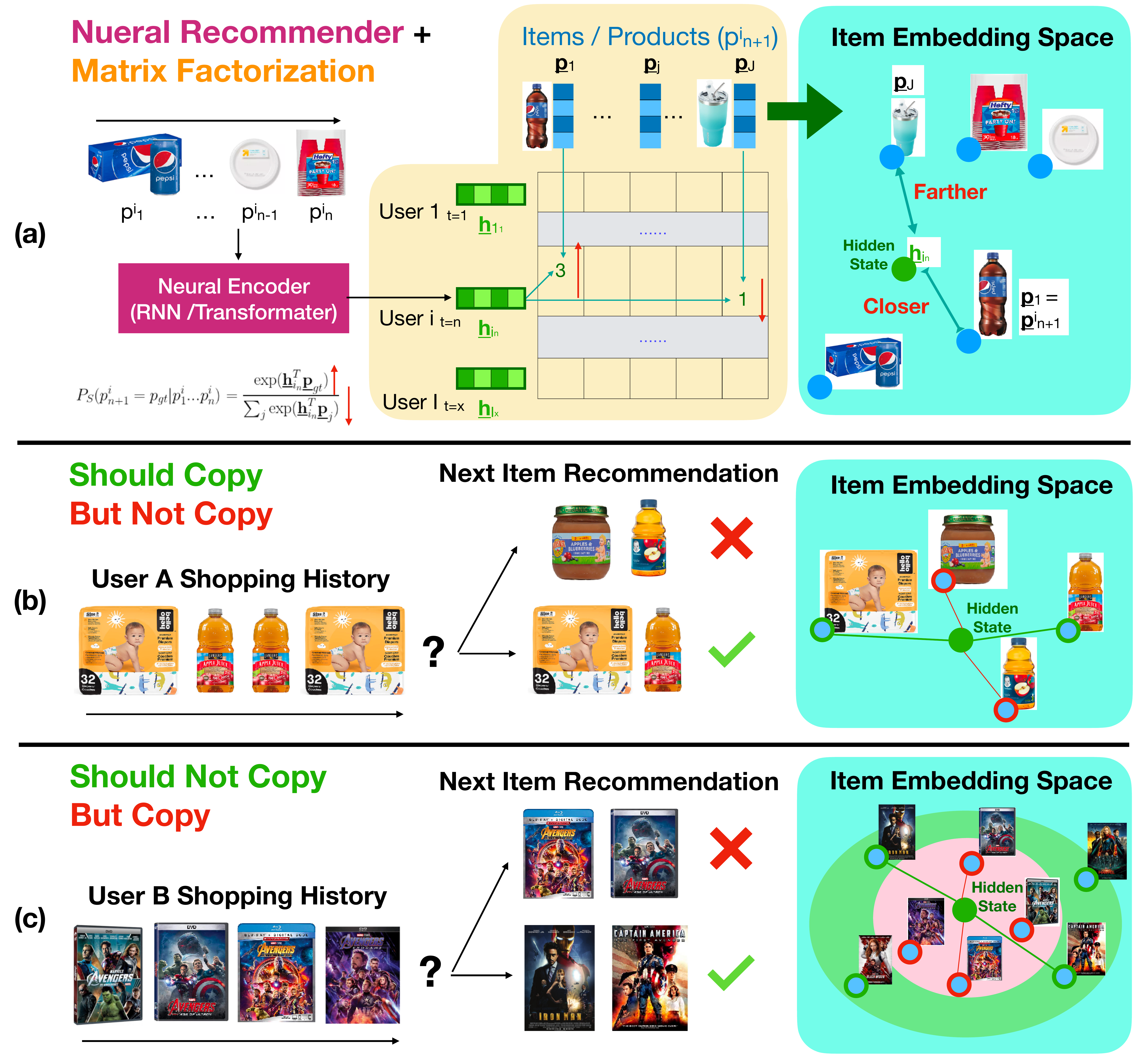}
\caption{The benefits and problems of the output softmax layer in a neural sequential recommender. (a) The output softmax layer implicitly factorizes the interaction matrix into the global item embeddings and the hidden states of the neural encoder. The similarity structure in the item embedding space helps recommender's generalization capability in this example.
%when user's multiple intents could be satisfied by a product. 
(b) In a dataset with many duplicated items, the recommender often needs to copy the items from the shopping history, but the item similarity structure in the embedding space does not allow the recommender to output the desired distribution. (c)  In a dataset with only few or no duplicated items, the model needs to learn not to recommend the items the users have already interacted with but to recommend something similar to them instead. The ideal distribution would form a donut shape in the item embedding space, which cannot be modeled by the single hidden state and static item embeddings in the softmax layer.}
\label{fig:softmax_all}
\end{figure*}

\begin{figure*}
\centering
\begin{subfigure}{.19\textwidth}
  \centering
  \includegraphics[width=1\linewidth]{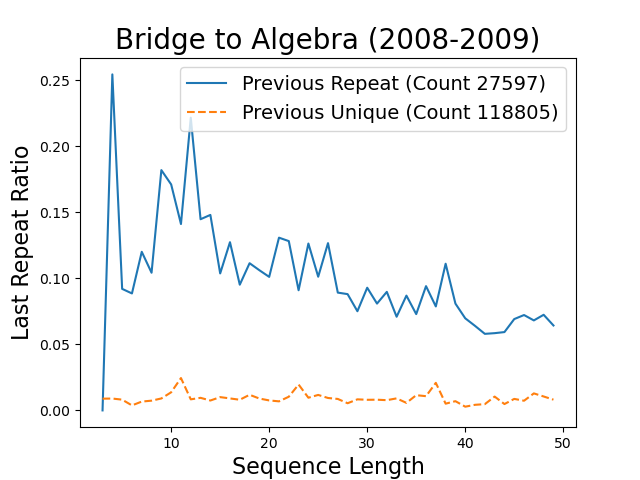}
  \caption{Math Exercises}
  \label{fig:algebra}
\end{subfigure}%
\begin{subfigure}{.19\textwidth}
  \centering
  \includegraphics[width=1\linewidth]{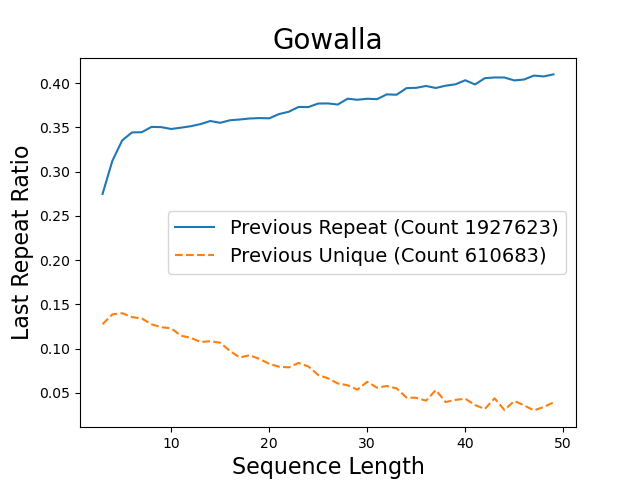}
  \caption{Locations}
  \label{fig:gowalla}
\end{subfigure}
\begin{subfigure}{.19\textwidth}
  \centering
  \includegraphics[width=1\linewidth]{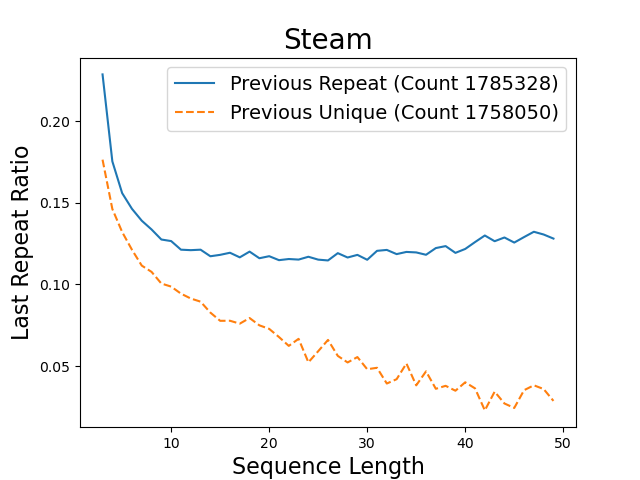}
  \caption{Video Games}
  \label{fig:steam}
\end{subfigure}
\begin{subfigure}{.19\textwidth}
  \centering
  \includegraphics[width=1\linewidth]{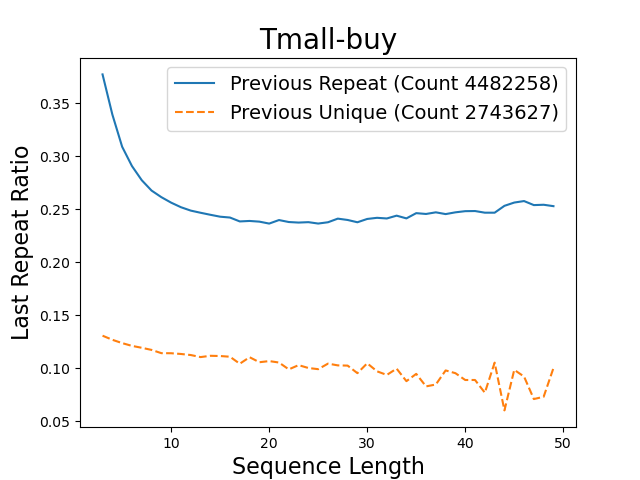}
  \caption{Purchases}
  \label{fig:tmall}
\end{subfigure}
\begin{subfigure}{.19\textwidth}
  \centering
  \includegraphics[width=1\linewidth]{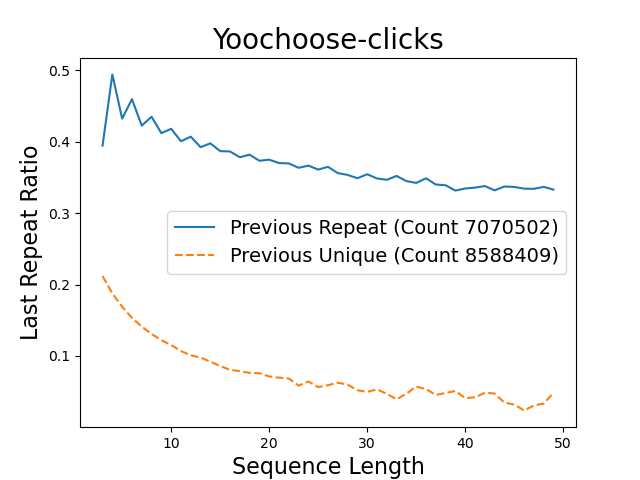}
  \caption{E-commerical Clicking}
  \label{fig:yoochoose}
\end{subfigure}
\caption{The probability of observing the repeated next item (i.e., the next item has already been in the input sequence) at the certain sequence length in x-axis. The blue curves are the probability if the input sequence has already had duplicated item(s), while the orange curves indicate the probability when every item in the input sequence is unique.}
%Previous repeat. Previous unique. Sequence Length. Last Repeat Ratio. }
\label{fig:repeat_stats}
\end{figure*}

%user's interaction history record.
%Sequence-aware recommender systems have various applications on 
%\citep{quadrana2018sequence}
%wide applications

Many recommendation tasks on the internet can be formulated as a sequential recommendation problem~\citep{quadrana2018sequence}, whose goal is to recommend the next item to each user based on the historical sequential interactions (e.g., click stream, purchasing record, and exercise practicing sequence) between the user and items~\citep{kang2018self}. 
In a sequential recommendation application, a good recommender often needs to capture the compositional meaning of multiple items in the input sequence, and many researchers have demonstrated that neural networks are able to model the complex interactions of the input items well and achieved state-of-the-art performances~\citep{wu2022survey}.

As shown in \autoref{fig:first_page}, a sequential recommender takes the item history of a user as the input and outputs a probability distribution of the next item. A list of the items with the highest predicted probabilities would be recommended to the user. The recommender can assign the highest probabilities to the items in the input history, with which the user interacted before. The repetition behavior is like copying the input items to the recommendation list. The recommender can also choose not to copy and encourage the user to explore the new items. \citet{li2023repetition} found that the modern neural recommender still cannot properly learn to copy or exclude the items from the history in many situations. Motivated by the practical need, we first identify that the output softmax layer, which is adopted by most of the state-of-the-art neural recommenders, is a major source of the problem and demonstrate a substantial performance improvement after alleviating the softmax problem.
%propose fixes to improve their performances.

%show that the difficulty could come from the 
%Most of the neural recommenders use an output softmax layer to compute the probability of the next item.

The softmax layer can be viewed as a matrix factorization layer \citep{yang2018breaking,rendle2020neural}. Instead of using a fixed user embedding as in a classic collaborative filtering method, the neural recommenders use a RNN (recurrent neural network) or transformer architecture to encode the historical input item sequence as a user embedding. Then, as shown in \autoref{fig:softmax_all} (a), the cross entropy loss and softmax layer encourage the high dot product between the generated user embedding and the embeddings of the possible next items. %and reduces the other dot products.

% \begin{figure}[t!]
% \centering
% \includegraphics[width=1\linewidth]{figs/matrix_factorization.pdf}
% \caption{}
% \label{fig:first_page}
% \end{figure}

% \begin{figure*}[t!]
% \centering
% \includegraphics[width=1\linewidth]{figs/softmax_limit.pdf}
% \caption{}
% \label{fig:softmax_limit}
% \end{figure*}

As in collaborative filtering, the softmax / matrix factorization layer has several benefits: It would encourage similar items to have similar item embeddings and similar input sequences to be encoded as similar user embeddings. In many cases, the similarity structure in the embedding space boosts the generalization capability of the system. We illustrate an example in \autoref{fig:softmax_all} (a): Assuming in our training data, we know that (i) many users like to buy a \textit{big bottle of drink} while buying \textit{party supplies} and (ii) users like to keep buying \textit{Pepsi}. Then, when a neural recommender sees a new user bought some \textit{Pepsi cans} and \textit{party supplies} before, it can output a user embedding that is between the \textit{Pepsi cans} and the average of \textit{2-liter drinks}. The user embedding would be close to the embeddings of a \textit{2-liter Pepsi bottle} because the \textit{2-liter Pepsi bottle} is similar to both other \textit{Pepsi} products and other \textit{2-liter drinks}.

Despite the effectiveness of softmax / matrix factorization layer, its item similarity structure is global and sometimes prevents the neural recommender from outputting the desired distribution. As shown in \autoref{fig:softmax_all} (b), if one user repeatedly bought a \textit{diaper} product and an \textit{apple juice} product, the next item should be very likely to be either the \textit{diaper} product or the \textit{apple juice} product. In order to output the distribution, the neural recommender needs to output a user embedding that is between the \textit{diaper} embedding and \textit{apple juice} embedding, which might be accidentally close to an embedding of \textit{apple-flavored baby food} and/or an embedding of \textit{apple juice for children}. In this case, the single hidden state embedding and static item embeddings in the softmax layer causes a difficulty in properly copying the items in the historical item sequence. This problem would become more serious when the embedding space is very crowded (e.g., the number of items is large).

On the other hand, the similarity structure could also force the neural recommender to improperly copy the previous items. As the example shown in \autoref{fig:softmax_all} (c), if one user just bought DVDs for the \textit{Avengers} movies, the users may want to watch the other \textit{Marvel} movies such as \textit{Iron Man} or \textit{Captain America} (i.e., something like \textit{Avengers} but not \textit{Avengers} themselves). However, a user embedding close to all the other \textit{Marvel} movies would be unavoidably close to the embeddings of \textit{Avengers} because the \textit{Avengers} movies are related to all the other \textit{Marvel} movies. This could force the neural recommender to keep recommending the movies that the user has seen before. In this case, the softmax layer causes difficulty in properly excluding the items in the historical item sequence.

To alleviate the issues caused by the output softmax layer, we adopt softmax-CPR~\citep{chang2023revisiting}, which is originally proposed to reduce the hallucinations of the language generation models. From \autoref{fig:softmax_all} (b), we can see that the main issue comes from the single hidden state in the item embedding space, so softmax-CPR uses different hidden states to compute the probabilities of different partitions of items. For example, we can use a hidden state for only the items in the input sequence (e.g., the \textit{apple juice} and \textit{diaper}) and another hidden state for the rest of the items (e.g., \textit{baby foods}). Then, the former hidden state could be placed between the \textit{apple juice} and \textit{diaper} without being interfered with by the other \textit{baby food} items.

We test our methods in RecBole~\citep{recbole[1.0],recbole[1.1.1]}. Compared to the softmax, which is used in most of the neural sequential recommenders, softmax-CPR improves 19\% on geometrically averaged NDCG@10 \citep{valcarce2020assessing} in 12 datasets and the improvement gaps are similar in SASRec~\citep{kang2018self} and GRU4Rec~\citep{hidasi2015session}. Our experiments also identify the source of improvement of RepeatNet~\citep{ren2019repeatnet} comes from alleviating the softmax issue rather than the self-attention mechanism. By identifying the source of the problem better, softmax-CPR can achieve larger improvements using a simpler model and easily be combined with any neural encoder.

%strong baseline

%each of them focus on a subset of products
%The idea is simple. We create some partitions .
%Then, we can still alleviate the softmax issues while leverage the generalization ability of global embeddings.
%unify

\subsection{Main Contributions}

\begin{itemize}[leftmargin=.2in,topsep=0pt]
\setlength\itemsep{0.0em}
\item We found that the single hidden state and static item embeddings in the output softmax layer prevent the neural sequential recommenders from learning the copying/excluding behavior of the users.
The perspective explains the improvement of RepeatNet~\citep{ren2019repeatnet} and simple post-processing~\citep{li2023repetition} in the datasets without duplicated items.
\item We adapt the softmax-CPR, which is recently proposed in \citet{chang2023revisiting} for NLP problems, for sequential recommendation tasks and implement them in RecBole.\footnote{Our code is released at \url{https://github.com/iesl/softmax_CPR_recommend}.} %Test the various softmax alternatives developed for NLP problems
\item Experiments in 12 datasets compare the various softmax alternatives unifiedly that might be able to solve the softmax bottleneck problems. We conduct detailed ablation studies to attribute the improvement to each modification we made in the softmax layer.

\end{itemize}

%% file: content/problems.tex
\section{Problems}
\label{sec:problems}
%In the introduction, we describe the problems using several examples. 
In this section, we will introduce the problems caused by the single hidden state and static item embeddings in the output softmax layer formally.  % Nikhil: 'doesn't like' sounds better than 'hate'..these two items are not similar 'for' this user.

\subsection{Duplicated Items}
In many applications, duplicated items appear frequently in the item sequence~\citep{bhagat2018buy,ariannezhad2022recanet}, and the repeated behavior could change in different domains, different experiment setups, or even just at different times in a sequence. For example, when a user buys a dress, the user often first explores lots of options (i.e., not many repetitions) and then repeatedly clicks a set of options to compare them. Once the user buys a dress, it is not very likely they will buy the same dress again, which means the repetition patterns for predicting the next click and next purchase are very different.

To show the diversity of the copying patterns, we plot the relationship between the repetition probability of the last item and the sequence length in \Cref{fig:repeat_stats}. The much higher blue curves than the orange curves imply that the copying probabilities drastically increase once the user starts to interact with items repeatedly. Moreover, the very different curves in different datasets suggest that it is very hard to manually design a general copy strategy that is applicable to various domains. 

%, we can see that 
%In \Cref{fig:repeat_stats}, we visualize the relation between the repetition probability of the last item and the sequence length. 
%show the statistics of those figures

%different domains often have repetition patterns. Even a small setup difference could completely change the copying behavior. 

%we are predicting the next movie the user clicks, 
%movie
%Browsing stage and decision stage
%Once a user starts to buy something again, the user might be more likely to buy it again and again.

\subsection{Softmax Bottleneck}
\label{sec:bottleneck}
The output probability from a softmax layer could be written as 
\begin{equation}
\label{eq:softmax}
P_{S}(p^i_{n+1}=p_x|i_n) = \frac{\text{exp}( \text{Logit}(x,i_n) )}{\sum_{j} \text{exp}( \text{Logit}(j,i_n) )} = \frac{\text{exp}(\vh_{i_n}^T \vp_x )}{\sum_{j} \text{exp}( \vh_{i_n}^T \vp_j )},
\end{equation}
where $i_n=p^i_{1}...p^i_{n}$ is the input item sequence with length 
$n$ from user $i$, $\vh_{i_n}$ is the hidden state for the sequence encoded by a neural encoder, and $\vp_x$ is the output item embeddings for item/product $x$. 
% Nikhil: the notations at the start of sentences are reducing the readability.

During training, the maximum likelihood estimation would increase the probability of observing the actual next product $p_{gt}$, $P_{S}(p^i_{n+1}=p_{gt}|i_n)$, by maximizing
$\vh_{i_n}^T \vp_{gt}$ and minimizing $\vh_{i_n}^T \vp_j$. That is, the hidden state $\vh_{i_n}$ would be pulled closer to $\vp_{gt}$ and pushed away from other item embeddings $\vp_j$. This implicit matrix factorization process is illustrated in \Cref{fig:softmax_all} (a). 

One main limitation of the softmax layer and matrix factorization is its static item embedding $\vp_{.}$. The meaning of each item is different for different users, but the item embeddings in the softmax layer are global and independent of the input item sequence. For example, most of users often buy items A and B together, so their global embeddings tend to be similar. However, if a kind of users kept only buying A because he/she likes A but doesn't like B, these two items are not similar to this kind of users at all. However, even though the users have repeatedly demonstrated their low probabilities of buying B, the recommender might be forced to keep recommending B by their similar global embeddings. As illustrated \Cref{fig:softmax_all} (b) and (c),  this discrepancy is especially serious when the user has a strong preference for copying or excluding the historical items.%interacting or not interacting with the historical items.

\citet{chang2022softmax} extend the concept of softmax bottleneck~\citep{yang2018breaking} and theoretically show that the structure of embeddings would create a multi-modal distribution that cannot be modeled by the single hidden state in the softmax layer (e.g., the bi-modal distribution in \Cref{fig:softmax_all} (b) and the donut shape distribution in (c)). Due to the latency and generalization concerns, we sometimes choose to share input and output embeddings and usually cannot use a very large hidden state size in an industrial recommender. Furthermore, we cannot observe all possible item distributions in the training data since every user could like different product combinations. For these reasons, it is hard for the softmax layer to reduce the multi-modal distribution by moving the item embeddings, which highlights the seriousness of the softmax bottleneck in many sequential recommendation tasks.
%These reasons make softmax layer hard to reduce the multi-modal distribution by moving the item embeddings, and thus make the softmax bottleneck especially serious in many sequential recommendation tasks. % Nikhil: For these reasons, it is hard for softmax layer to reduce the multi-modal distribution by moving the item embeddings which highlights the seriousness of softmax bottlneck, especially in many sequential recommendation tasks.

%often unable to alleviate the problem by rearrange .

%During training, we usually cannot 
%the copy behavior creates multi-modal distribution in the embedding space
%softmax cannot model such distribution
%dimensionality

%subspace
%cite my theoretical paper
%bottleneck~\citep{yang2018breaking,chang2022softmax}

%we cannot observe all possible item distributions in the input
%(every user could like different kind of product combination)
%global embedding is limited

%In many industrial recommendation system
%we have strict requirement on latency and computational cost
%so the hidden state dimension is small
%the Softmax Bottleneck would be serious

%softmax formula
%Cp some figures from unity to explain that different datasets have different copy probability

%, the item embedding.
%From the matrix factorization perspective, the problem 
%global embedding
%local meaning

%copy behavior is common

%Why single embedding is not enough
%The connection between repeated items and softmax bottleneck is not very clear

%% file: content/methods.tex
\section{Solutions}

In this section, we introduce several potential methods that could alleviate the softmax bottleneck problems.

% and
\subsection{Post Processing}

In industry, the repetition issues are often alleviated by business insights and heuristic rules. For example, a product manager might notice that most users won't watch the same movie twice and many movies in the recommendation list have been watched, so we could simply add a post-processing step to remove the movies that have been watched from our recommendation candidate list~\citep{li2023repetition}. Nevertheless, the simple statistics in \Cref{fig:repeat_stats} suggest that finding generally applicable rules is difficult. Moreover, in some domains, the rules might quickly become too complicated to manage. For example, a user might want a food delivery app to recommend something new but sometimes prefers to buy food from the restaurants he/she has tried more than $m$ times (to exclude the restaurants that are too bad to try again). In these cases, learning the copy patterns from a large training set should be an easier and more effective approach than the manually designed heuristic rules. 
%diverse copying behavior
% Nikhil: suggests that finding generally applicable rules is difficult..

%CPR represents three methods, context partition, pointer network, and reranker partition.

\begin{figure}[t!]
\centering
\includegraphics[width=1\linewidth]{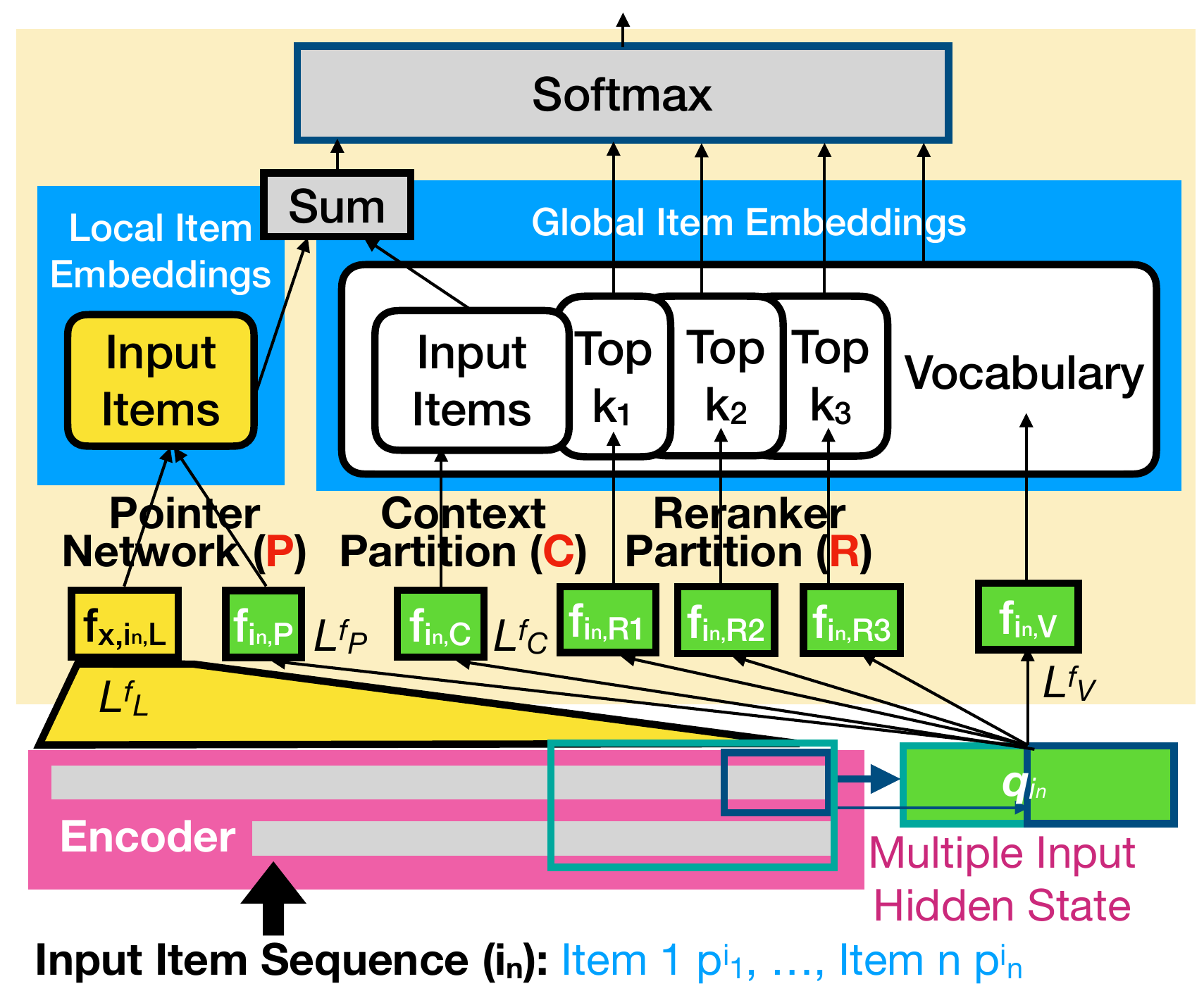}
\caption{The architecture of softmax-CPR and multiple input hidden states (Mi).}
\label{fig:softmax_cpr}
\end{figure}

%and $\vf_{i_n,V}$ is the hidden state for the other products such as baby foods. Then, . 

\subsection{Softmax-CPR}
In this section, we briefly review Softmax-CPR proposed by \citet{chang2023revisiting}. If you would like to know more details of its motivation and formulation, please refer to  \citet{chang2023revisiting} or \Cref{sec:cpr_details}. 

Softmax-CPR combines three methods: \textbf{C}ontext partition, \textbf{P}ointer network, and \textbf{R}eranker partition, to improve the output softmax layer. First, we introduce the context partition method. Context partition makes a small change in the logit computation in the softmax layer. In \Cref{eq:softmax}, the softmax layer lets $\text{Logit}(x,i_n) = \vh_{i_n}^T \vp_x$. In context partition, the logit of item $x$ 
\begin{equation}
\label{eq:logit_C}
\text{Logit}_{C}(x,i_n)=\left\{
\begin{matrix*}[l]
\vf_{i_n,C}^T \vp_x \;\; \text{if} \; x \in i_n\\[3pt]
\vf_{i_n,V}^T \vp_x \;\; \text{O/W}
\end{matrix*}\right.,
\end{equation}
where $\vf_{i_n,C}=L_C^f(\vh_{i_n})$ and $\vf_{i_n,V}=L_V^f(\vh_{i_n})$ are the linear projections of the hidden state. In this paper, $L_{.}^f(\vh_{i_n}) = W_{.}\vh_{i_n} + \vb_{.}$ (e.g., $L_{C}^f(\vh_{i_n}) = W_{C}\vh_{i_n} + \vb_{C}$) and each linear projection layer would learn different parameters (weights $W_{.}$ and bias $\vb_{.}$) during training. 

The context partition allows the recommender to learn when to copy input items and when to exclude the input items. For example, in \Cref{fig:softmax_all} (b), the recommender can place $\vf_{i_n,C}$ between the \textit{apple juice} and the \textit{diaper} without being interfered by the \textit{baby foods} because $\vf_{i_n,C}$ is the hidden state only for the input items.
Similarly, in \Cref{fig:softmax_all} (c), the recommender can learn to output a very small value of $\vf_{i_n,C}^T \vp_x$ to exclude all the previously seen movies, while placing $\vf_{i_n,V}$ at the center of the movies we should recommend.

The context partition is related to a pointer network. The main difference is that the pointer network computes the logit of the input items by $\vf_{i_n,P}^T \vf_{x,i_n,L}$ instead of $\vf_{i_n,P}^T \vp_x$, where $\vf_{i_n,P}=L_{P}^f(\vh_{i_n})$ and $\vf_{x,i_n,L}$ is the average of a linearly projected hidden state embedding corresponding to the item $x$ (see \Cref{eq:local_emb} in the appendix for its formula). This means that the pointer network allows the recommender to predict the local and context-dependent embedding for the input items.

In the context partition or pointer network, 
%One limitation of the context partition and pointer network is that 
we only compute the logits of input items separately using projected hidden states. However, some unexplored items are also likely to appear next and they might also encounter the softmax bottleneck problem. To alleviate the issue, the reranker partition computes the logits of the most likely $k$ items separately using another new hidden state (see the term $\vf_{i_n,R1}^T \vp_x$ below for an example). 

As illustrated in \Cref{fig:softmax_cpr}, Softmax-CPR combines all the above methods by
\begin{equation}
\label{eq:logit_CPR}
\text{Logit}_{CPR}(x,i_n)=
\left\{
\begin{matrix*}[l]
\vf_{i_n,C}^T \vp_x + \vf_{i_n,P}^T \vf_{x,i_n,L} \;\; \text{if} \; x \in i_n\\[3pt] 
\vf_{i_n,R1}^T \vp_x \;\; \text{if} \; x \in P(k_1) - i_n \\[3pt]
\vf_{i_n,R2}^T \vp_x \;\; \text{if} \; x \in P(k_2) - P(k_1) - i_n \\[3pt]
\vf_{i_n,R3}^T \vp_x \;\; \text{if} \; x \in P(k_3) - P(k_2) - P(k_1) - i_n \\[3pt]
\vf_{i_n,V}^T \vp_x \;\; \text{O/W}
\end{matrix*}\right.,
\end{equation}
where $P(k_3)$ is the top $k_3$ items with the highest $\vf_{i_n,V}^T \vp_x$, $P(k_2)$ is the top $k_2$ words with the highest logits, similarly for $P(k_1)$, and $\vf_{i_n,Ry}=L_{Ry}^f(\vh_{i_n})$. Note that this method could be easily combined with maximum inner product search~\citep{bruch2023bridging}. We can just use $\vf_{i_n,V}$ to search the possible next items and use other hidden states to adjust the logits of these possible next items and input items. Since the number of input items $|i_n|$ and $k_3$ are much smaller than the total number of items, the computational overhead should be relatively small.

%learn when to copy and when not to copy
%~\citep{chang2023revisiting}
%merging in the logit level to exclude the items in the history
%When introducing softmax cpr, mentioning hallucination (should copy but not copy) and repetition (should not copy but copy)

%For the methods that use different $S$, we can simply determine an order of computing the dot products and let the later dot products overwrite the existing values. In our experiments, we always use the order illustrated in \autoref{fig:all_partition}. That is, we compute the logits ($\text{Logit}_{CPR}(x,i_n)$) by

\subsection{Multiple Input Hidden States (Mi)}
\label{sec:Mi}
%\citet{chang2022softmax} show that 

All the new hidden states $\vf_{i_n,.}$ are the projection of $\vh_{i_n}$. The limited dimension of $\vh_{i_n}$ would force $\vf_{i_n,.}$ to linearly depend on each other and not be able to move freely in the item embedding space. To solve this issue, \citet{chang2022softmax} concatenate multiple input hidden states $\vh_{i_n}$ and project them into a new hidden state $\vq_{i_n}$ to expand its dimensionality. We combine their method with softmax-CPR in \Cref{fig:softmax_cpr} by replacing $\vh_{i_n}$ with $\vq_{i_n}$ in \Cref{eq:logit_CPR}.
%by using a 
%to prevent the .

%create the dependency of $\vf_{i_n,.}$. 
%\citep{chang2022softmax}
%The multiple hidden states would share some dependency if they are projected by $\vh_{i_n}$. new bottleneck
%all the hidden states would have some dependency
%Not using equation?

\subsection{Mixture of Softmax (MoS)}

In \Cref{fig:softmax_all} (b), one possible solution is to put one hidden state near the \textit{diaper} and another hidden state near the \textit{apple juice} to model its multi-modal distribution. The mixture of softmax (MoS) is proposed to achieve the goal~\citep{yang2018breaking,lin2021breaking,chang2022softmax}. MoS and softmax-CPR both use multiple hidden states, while their roles of each hidden state are different. In MoS, we need to compute the dot product between every hidden state and all the item embeddings; in softmax-CPR, we partition the item set and each hidden state only determines the logits/probabilities of the items in a partition (e.g., only the input items in the context partition). Compared to softmax-CPR, MoS is more computationally expensive and does not explicitly model users' repetition behavior.

%The main difference between mixture of softmax (MoS)
%multiple hidden state
%In MoS, we use multiple hidden states 
%\citep{lin2021breaking}
%expensive because we need to do 

\subsection{RepeatNet}

Inspired by CopyNet~\citep{gu2016incorporating} / pointer network, RepeatNet~\citep{ren2019repeatnet} explicitly models the probability of copying the items from the input. In their paper, they do not test its performance on the datasets without duplicated items. Compared to softmax-CPR, RepeatNet has several disadvantages. First, RepeatNet introduces many extra parameters to GRU4Rec, which increases the computational overhead and the difficulties in identifying the source of the improvement. Second, when computing the probabilities of copying the items, RepeatNet does not leverage the global item similarity structure, which might hurt the generality of the model (see \Cref{fig:softmax_all} (a) for an example). Third, RepeatNet does not solve the softmax bottleneck problem for the items that are not in the input sequence.

%However, they 
%We could also explicitly model the probabilities of copying the items from the input sequence. 
%One representative work is RepeatNet~\citep{ren2019repeatnet}, which uses a pointer network / copying mechanism / self attention to improve prediction on duplicated items. 

%extra parameters
%introduce unnecessary computation

%further improve RepeatNet in the sequential recommendation tasks. Compared to RepeatNet, softmax-CPR prevents the similarity structure of the item embedding from interfering the distribution of the likely next items while leveraging the similarity structure to improve the distribution of the items in the input sequence.

%% file: content/experiments.tex
% \usepackage{multirow}
\begin{table}[]
\scalebox{0.7}{
\begin{tabular}{lll|ccc|cc}
& & & \multicolumn{3}{c|}{Dataset size (k)} & \multicolumn{2}{c}{Config} \\
\multicolumn{2}{c}{Dataset}                         & Item Type                 & \#User & \#Item & \# Inter & $|\vh_{i_n}|$ & bsz \\ \hline
\multirow{3}{*}{Amazon-2014~\citep{mcauley2015image}}      & Beauty          & \multirow{3}{*}{Products} & 1210                         & 249                          & 2023                                & 64                      & {[}64,128{]}            \\
                                  & Books            &                           & 8026                         & 2330                         & 22507                               & 32                      & 32                      \\
                                  & Video Games      &                           & 827                          & 50                           & 1325                                & 64                      & {[}64,128{]}            \\
\multirow{2}{*}{Movie Lens~\citep{harper2015movielens}}       & 10m             & \multirow{2}{*}{Movies}   & 70                           & 11                           & 10000                               & 64                      & 128                     \\
                                  & 1m              &                           & 6                            & 4                            & 1000                                & 64                      & {[}64,128{]}            \\
\multicolumn{2}{l}{Twitch-100k~\citep{rappaz2021recommendation}}                     & Videos                    & 100                          & 740                          & 3052                                & 48                      & 32                      \\
\multicolumn{2}{l}{Yelp-2018\tablefootnote{\url{https://www.yelp.com/dataset}}}                       & Stores                    & 1326                         & 175                          & 5262                                & 48                      & 32                      \\
\multicolumn{2}{l}{Bridge to Algebra (2008-2009)~\citep{stamper2010challenge}} & Exercises                  & 3                            & 1259                         & 8918                                & 48                      & 32                      \\
\multicolumn{2}{l}{Gowalla~\citep{cho2011friendship}}                         & Locations                  & 107                          & 1281                         & 6443                                & 32                      & 32                      \\
\multicolumn{2}{l}{Steam~\citep{kang2018self}}                           & Games                      & 2568                         & 32                           & 7793                                & 64                      & {[}64,128{]}            \\
\multicolumn{2}{l}{Tmall-buy~\citep{Tmall}}                       & Products                  & 886                          & 1144                         & 9349                                & 32                      & 32                      \\
\multicolumn{2}{l}{Yoochoose-clicks~\citep{ben2015recsys}}                & Products                  & 9250                         & 53                           & 33004                               & 64                      & 128                    
\end{tabular}
}
\caption{The dataset sizes are reported by the number of thousands (k). We adjust the hidden state size ($|\vh_{i_n}|$) and batch size (bsz) accordingly under our GPU memory constraint.}
%to fit the models into our GPU memory. }
\label{tb:stats}
\end{table}

% Amazon 2014~\citep{mcauley2015image}
% Movie Lens 1m and 10m~\citep{harper2015movielens}

% Twitch-100k~\citep{rappaz2021recommendation}
% Yelp-2018\footnote{\url{https://www.yelp.com/dataset}}
% Algebra I (2008-2009)~\citep{stamper2010challenge}
% Gowalla~\citep{cho2011friendship}
% Steam~\citep{kang2018self}
% Tmall-buy~\citep{Tmall}
% Yoochoose-clicks~\citep{ben2015recsys}

% \usepackage{multirow}
\begin{table*}[t!]
\centering
\scalebox{0.76}{
% \usepackage{multirow}
\begin{tabular}{cl|cccccccccccccc|}
                   &                               & \multicolumn{6}{c|}{Amazon-2014}                                                                    & \multicolumn{4}{c|}{MovieLens}                                     & \multicolumn{2}{c|}{\multirow{2}{*}{Twitch-100k}} & \multicolumn{2}{c|}{\multirow{2}{*}{Yelp-2018}}  \\
\multicolumn{1}{l}{}                   &                               & \multicolumn{2}{c|}{Beauty}    & \multicolumn{2}{c|}{Books}      & \multicolumn{2}{c|}{Video Games}      & \multicolumn{2}{c|}{10m}         & \multicolumn{2}{c|}{1m}          & \multicolumn{2}{c|}{}                        & \multicolumn{2}{c|}{}                                             \\
\multicolumn{1}{l}{}                   &                               & NDCG          & \multicolumn{1}{c|}{HR}           & NDCG          & \multicolumn{1}{c|}{HR}           & NDCG          & \multicolumn{1}{c|}{HR}           & NDCG           & \multicolumn{1}{c|}{HR}            & NDCG           & \multicolumn{1}{c|}{HR}            & NDCG                 & \multicolumn{1}{c|}{HR}                  & NDCG                & \multicolumn{1}{c|}{HR}\\ \hline
\multirow{9}{*}{\textbf{SASRec}}       & Softmax                       & 1.16          & 2.19          & 3.30          & 5.81          & 4.12          & 7.97          & 15.72          & 26.67          & 16.75          & 29.45          & 8.41                 & 15.51                & 1.66                & 3.36                \\
                                       & Softmax + Mi                  & 1.18          & 2.20          & 3.23          & 5.77          & 3.79          & 7.48          & 15.80          & 26.69          & 16.67          & 29.06          & 8.08                 & 15.03                & 1.67                & 3.36                \\
                                       & Softmax + C                   & 1.41          & 2.41          & 3.83          & 6.46          & 4.41          & 8.27          & 19.12          & 31.13          & 20.70          & 34.19          & 9.14                 & 16.39                & 1.94                & 3.82                \\
                                       & Softmax + CP                  & \textbf{1.45} & \textbf{2.52} & 3.94          & 6.71          & 4.54          & 8.59          & 18.62          & 30.51          & 20.69          & \textbf{34.67} & \textbf{9.45}        & \textbf{16.93}       & 2.04                & 3.91                \\
                                       & Softmax + CPR:100             & 1.38          & 2.42          & 4.15          & 6.89          & 4.57          & \textbf{8.69} & \textbf{19.32} & \textbf{31.32} & 20.79          & 34.25          & 9.11                 & 15.94                & \textbf{2.22}       & 4.24                \\
                                       & Softmax + CPR:100 + Mi        & 1.37          & 2.41          & \textbf{4.30} & \textbf{7.20} & \textbf{4.47} & 8.40          & 18.90          & 30.73          & \textbf{20.82} & 34.49          & 9.06                 & 15.91                & 2.21                & 4.24                \\
                                       & Softmax + CPR:20,100,500 + Mi & 1.39          & 2.43          & 3.93          & 6.60          & 4.46          & 8.58          & 19.19          & 30.93          & 20.48          & 33.61          & 8.58                 & 14.88                & 2.20                & \textbf{4.27}       \\
                                       & Mixture of Softmax (MoS) & 1.19          & 2.24          & 3.24          & 5.75          & 3.74          & 7.35          & 15.88          & 26.82          & 17.05          & 29.83          & 8.17                 & 15.19                & 1.69                & 3.42                \\
                                       & Softmax w/o Duplication~\citep{li2023repetition}       & 1.34          & 2.42          & 3.73          & 6.27          & 4.42          & 8.35          & 18.35          & 30.19          & 20.06          & 33.81          & 9.01                 & 16.13                & 1.85                & 3.64                \\ \hline
\multirow{9}{*}{\textbf{GRU4Rec}}      & Softmax                       & 1.43          & 2.67          & 3.09          & 5.70          & 4.45          & 8.64          & 14.19          & 24.17          & 16.05          & 28.03          & 8.36                 & 15.55                & 1.68                & 3.42                \\
                                       & Softmax + Mi                  & 1.47          & 2.69          & 3.30          & 5.92          & 4.58          & 8.79          & 14.58          & 25.04          & 16.55          & 28.94          & 8.03                 & 14.98                & 1.76                & 3.52                \\
                                       & Softmax + C                   & 1.59          & 2.88          & 3.97          & 6.66          & 4.95          & 9.36          & 17.78          & 29.24          & 20.01          & 32.86          & 9.25                 & 16.50                & 2.02                & 3.92                \\
                                       & Softmax + CP                  & 1.61          & 2.94          & 4.07          & 6.83          & \textbf{5.10} & 9.41          & 17.46          & 28.64          & 19.63          & 32.91          & 9.14                 & 16.09                & 2.00                & 3.85                \\
                                       & Softmax + CPR:100             & 1.78          & \textbf{3.22} & 4.28          & 7.06          & 5.05          & \textbf{9.49} & 17.78          & 29.01          & 20.35          & 33.73          & 9.04                 & 15.82                & 2.27                & 4.35                \\
                                       & Softmax + CPR:100 + Mi        & 1.72          & 3.15          & \textbf{4.42} & \textbf{7.23} & 5.07          & 9.43          & \textbf{18.09} & \textbf{29.43} & \textbf{21.00} & \textbf{34.52}          & \textbf{9.32}                 & \textbf{16.20}                & \textbf{2.37}                & \textbf{4.51}                \\
                                       & Softmax + CPR:20,100,500 + Mi & \textbf{1.73} & 3.11          & 4.37          & 7.14          & 5.02          & 9.33          & 17.87          & 29.09          & 20.44          & 33.63          & 8.80                 & 15.20                & 2.31                & 4.39                \\
                                       & Mixture of Softmax (MoS)                           & 1.46          & 2.73          & 3.15          & 5.76          & 4.06          & 8.00          & 14.40          & 24.50          & 16.14          & 28.06          & 7.90                 & 14.69                & 1.73                & 3.50                \\
                                       & Softmax w/o Duplication~\citep{li2023repetition}       & 1.60          & 2.91          & 3.71          & 6.26          & 4.83          & 9.09          & 16.85          & 27.68          & 18.54          & 31.72          & 8.94                 & 16.03                & 1.94                & 3.80                \\ \hline
\multicolumn{1}{l}{\textbf{RepeatNet}} & -                       & 1.75          & 2.88          & 3.94          & 6.36          & 4.47          & 8.36          & 18.09          & 29.20          & 18.71          & 31.08          & 8.52                 & 14.91                & 2.02                & 3.88               

\end{tabular}
}
\caption{We compare the test performance (\%) of NDCG@10 and HR@10 in 7 datasets without duplicated items. C, P, R means context partition, pointer network, and reranker partition, respectively. 20,100,500 refers to $k_1=20$, $k_2=100$ and $k_3=500$; Mi means the multiple input hidden state enhancement. The best values given the same neural encoder are highlighted.}
\label{tb:nodup_dataset}
\end{table*}
%Please see \autoref{eq:logit_CPR} for the details of CPR.

\begin{table*}[t!]
\centering
\scalebox{0.8}{
\begin{tabular}{cl|cccccccccc|}
\multicolumn{1}{l}{}                   &                               & \multicolumn{2}{c|}{Bridge to Algebra}     & \multicolumn{2}{c|}{Gowalla}     & \multicolumn{2}{c|}{Steam}       & \multicolumn{2}{c|}{Tmall-buy}       & \multicolumn{2}{c|}{Yoochoose-clicks}     \\
\multicolumn{1}{l}{}                   &                               & NDCG           & \multicolumn{1}{c|}{HR}            & NDCG           & \multicolumn{1}{c|}{HR}            & NDCG           & \multicolumn{1}{c|}{HR}            & NDCG           & \multicolumn{1}{c|}{HR}            & NDCG           & \multicolumn{1}{c|}{HR}             \\ \hline
\multirow{9}{*}{\textbf{SASRec}}       & Softmax                       & 85.66          & 90.42          & 29.28          & 40.39          & 15.67          & 20.28          & 22.44          & 26.60          & 35.74          & 57.28          \\
                                       & Softmax + Mi                  & 85.68          & 89.72          & 29.72          & 40.72          & 15.77          & 20.47          & 22.64          & 26.80          & 36.62          & 57.93          \\
                                       & Softmax + C                   & 86.25          & 91.15          & 32.23          & 45.15          & 16.32          & 21.13          & 25.29          & 30.36          & 37.26          & 58.93          \\
                                       & Softmax + CP                  & 85.60          & 89.75          & 32.88          & 45.68          & 16.30          & 21.05          & 25.58          & 30.50          & 37.43          & 59.02          \\
                                       & Softmax + CPR:100             & 87.40          & 91.09          & 33.03          & 46.17          & 16.43          & 21.31          & 25.73          & \textbf{30.70} & 37.79          & 59.15          \\
                                       & Softmax + CPR:100 + Mi        & 88.19          & \textbf{92.19} & 33.41          & 46.29          & \textbf{16.48} & \textbf{21.39} & \textbf{25.74} & 30.58          & 39.03          & \textbf{59.69} \\
                                       & Softmax + CPR:20,100,500 + Mi & \textbf{88.81} & 92.07          & \textbf{33.92} & \textbf{46.64} & 16.34          & 21.15          & 25.58          & 30.22          & \textbf{39.26} & 59.68          \\
                                       & Mixture of Softmax (MoS)                           & 84.77          & 89.78          & 29.74          & 40.87          & 15.90          & 20.49          & 23.07          & 27.28          & 35.59          & 57.07          \\
                                       & Softmax w/o Duplication~\citep{li2023repetition}       & 80.13          & 82.89          & 3.92           & 7.00           & 4.89           & 9.15           & 4.29           & 6.28           & 17.00          & 27.84          \\ \hline
\multirow{9}{*}{\textbf{GRU4Rec}}      & Softmax                       & 85.10          & 89.23          & 28.37          & 39.48          & 15.35          & 19.88          & 22.06          & 26.42          & 36.19          & 56.97          \\ 
                                       & Softmax + Mi                  & 84.68          & 89.01          & 27.99          & 39.06          & 15.69          & 20.26          & 21.76          & 26.05          & 36.39          & 57.15          \\
                                       & Softmax + C                   & 85.86          & 89.75          & 32.23          & 45.18          & 16.29          & 21.04          & 25.18          & 30.25          & 37.46          & 58.54          \\
                                       & Softmax + CP                  & 86.24          & 91.06          & 32.48          & 45.43          & 16.32          & 21.06          & 25.45          & 30.36          & 37.90          & 58.76          \\
                                       & Softmax + CPR:100             & 88.56          & \textbf{92.35} & 33.01          & 46.08          & 16.36          & 21.15          & \textbf{25.77} & 30.34          & 38.35          & 59.15          \\
                                       & Softmax + CPR:100 + Mi        & 88.81          & 92.19          & \textbf{33.22} & \textbf{46.09} & \textbf{16.49} & \textbf{21.35} & 25.54          & 30.01          & \textbf{38.72} & \textbf{59.42} \\
                                       & Softmax + CPR:20,100,500 + Mi & \textbf{89.46} & 92.29          & 33.18          & 45.93          & 16.41          & 21.19          & 25.72          & \textbf{30.43} & 38.54          & 59.20          \\
                                       & Mixture of Softmax (MoS)                           & 86.11          & 90.30          & 27.91          & 38.60          & 15.89          & 20.41          & 21.50          & 25.75          & 36.39          & 56.97          \\
                                       & Softmax w/o Duplication~\citep{li2023repetition}       & 79.06          & 81.67          & 3.93           & 7.05           & 4.65           & 8.72           & 4.24           & 6.32           & 16.80          & 27.44          \\ \hline
\multicolumn{1}{l}{\textbf{RepeatNet}} & - & 77.44          & 81.70          & 33.83          & 45.88          & 16.28          & 20.90          & 25.67          & 30.17          & 38.00          & 58.53                    

\end{tabular}
}
\caption{The test performance (\%) in 5 datasets with duplicated items. The notations are the same as \Cref{tb:nodup_dataset}.}
\label{tb:dup_dataset}
\end{table*}

% \usepackage{multirow}
\begin{table*}[t!]
\centering
\scalebox{0.85}{
\begin{tabular}{cl|c|cc|ccccccccc|}
\multicolumn{1}{l}{}                   &                               & Model & Training & Testing & \multicolumn{9}{c|}{Geometric Mean}                                                                                                   \\
 &                               & Size (M)                            & Time (s) & Time (s)                              & \multicolumn{3}{c|}{7 datasets w/o dup.} & \multicolumn{3}{c|}{5 datasets w/ dup.} & \multicolumn{3}{c|}{All 12 datasets} \\
                  &                          &     &                      &                               & NDCG                  & HR       & \multicolumn{1}{c|}{MRR}              & NDCG                  & HR              & \multicolumn{1}{c|}{MRR}     & NDCG              & HR           &MRR   \\ \hline
\multirow{9}{*}{\textbf{SASRec}}       & Softmax                       & 5.34                                            & 1178.95                                           & 22.43                                             & 4.79          & 8.82           & 3.57          & 31.60          & 40.78          & 28.56          & 10.75          & 17.00          & 8.70           \\
                                       & Softmax + Mi                  & 5.36                                            & 1306.67                                           & 23.71                                             & 4.71          & 8.68           & 3.50          & 31.95          & 41.02          & 28.95          & 10.70          & 16.89          & 8.66           \\
                                       & Softmax + C                   & 5.35                                            & 1535.21                                           & 28.53                                             & 5.57          & 9.78           & 4.29          & 33.59          & 43.49          & 30.32          & 12.04          & 18.54          & 9.92           \\
                                       & Softmax + CP                  & 5.36                                            & 1601.77                                           & 29.13                                             & 5.69          & 10.02          & 4.36          & 33.77          & 43.48          & 30.56          & 12.20          & 18.80          & 10.04          \\
                                       & Softmax + CPR:100             & 5.37                                            & 2013.69                                           & 34.43                                             & 5.77          & \textbf{10.07} & \textbf{4.45} & 34.10          & 43.88          & 30.86          & 12.35          & 18.92          & 10.20          \\
                                       & Softmax + CPR:100 + Mi        & 5.41                                            & 2213.77                                           & 35.64                                             & \textbf{5.75} & 10.06          & 4.44          & 34.49          & \textbf{44.09} & 31.32          & \textbf{12.39} & \textbf{18.95} & \textbf{10.25} \\
                                       & Softmax + CPR:20,100,500 + Mi & 5.41                                            & 2999.23                                           & 47.63                                             & 5.64          & 9.86           & 4.34          & \textbf{34.58} & 43.94          & \textbf{31.48} & 12.26          & 18.71          & 10.14          \\
                                       & Mixture of Softmax (MoS)                           & 5.35                                            & 1779.93                                           & 33.78                                             & 4.74          & 8.75           & 3.52          & 31.88          & 41.08          & 28.82          & 10.73          & 16.98          & 8.67           \\
                                       & Softmax w/o Duplication~\citep{li2023repetition}       & 5.34                                            & 1183.55                                           & 22.61                                             & 5.41          & 9.61           & 4.12          & 10.23          & 15.61          & 8.44           & 7.11           & 11.84          & 5.61           \\ \hline
\multirow{9}{*}{\textbf{GRU4Rec}}      & Softmax                       & 5.42                                            & 1842.56                                           & 31.24                                             & 4.85          & 8.99           & 3.58          & 31.20          & 40.23          & 28.22          & 10.77          & 17.09          & 8.68           \\
                                       & Softmax + Mi                  & 5.45                                            & 2239.74                                           & 33.37                                             & 4.98          & 9.14           & 3.71          & 31.18          & 40.19          & 28.20          & 10.93          & 17.25          & 8.85           \\
                                       & Softmax + C                   & 5.42                                            & 2333.21                                           & 37.84                                             & 5.75          & 10.16          & 4.39          & 33.55          & 43.24          & 30.36          & 12.24          & 18.90          & 10.06          \\
                                       & Softmax + CP                  & 5.43                                            & 2355.90                                           & 38.45                                             & 5.76          & 10.14          & 4.41          & 33.80          & 43.48          & 30.60          & 12.29          & 18.93          & 10.12          \\
                                       & Softmax + CPR:100             & 5.44                                            & 3103.13                                           & 43.18                                             & 6.02          & 10.55          & 4.63          & 34.27          & 43.82          & 31.11          & 12.68          & 19.42          & 10.47          \\
                                       & Softmax + CPR:100 + Mi        & 5.50                                            & 3480.22                                           & 45.60                                             & \textbf{6.13} & \textbf{10.69} & \textbf{4.72} & 34.39          & \textbf{43.83} & 31.27          & \textbf{12.83} & \textbf{19.57} & \textbf{10.62} \\
                                       & Softmax + CPR:20,100,500 + Mi & 5.50                                            & 4128.41                                           & 57.51                                             & 6.01          & 10.44          & 4.64          & \textbf{34.42} & \textbf{43.83} & \textbf{31.30} & 12.69          & 19.31          & 10.52          \\
                                       & Mixture of Softmax (MoS)                           & 5.43                                            & 2448.34                                           & 42.67                                             & 4.81          & 8.91           & 3.56          & 31.27          & 40.15          & 28.33          & 10.72          & 16.98          & 8.66           \\
                                       & Softmax w/o Duplication~\citep{li2023repetition}       & 5.42                                            & 1798.91                                           & 31.00                                             & 5.52          & 9.83           & 4.20          & 10.06          & 15.42          & 8.27           & 7.14           & 11.92          & 5.61           \\ \hline
\multicolumn{1}{l}{\textbf{RepeatNet}} & -                       & 15.97                                           & NA                                                & NA                                                & 5.63          & 9.69           & 4.39          & 33.41          & 42.48          & 30.42          & 12.08          & 18.26          & 10.06         
\end{tabular}
}
\caption{We report the model size and average time of training/testing the models on Amazon-2014 Books for 1 epoch. The other notations are the same as \Cref{tb:nodup_dataset}. }
\label{tb:all_datasets}
%for getting the probabilities of next sequence in the validation set. }
\end{table*}

%RepeatNet~\citep{ren2019repeatnet}
%\subsection{Datasets}

\section{Experiments}
All the experiments are done in RecBole~\citep{recbole[1.0],recbole[1.1.1]}, a library that provides various recommendation models and datasets. 
We select 12 datasets from RecBole that are large enough and widely used in the previous work to make the results more representative and less sensitive to the hyperparameter setup and random seeds~\citep{ferrari2021troubling}. The datasets come from various domains and have various sizes. We report their statistics in \Cref{tb:stats}.  

%Separating duplicated and not duplicated datasets
%The standard output softmax layer used 

\subsection{Models and Baselines}
We implement the following softmax alternatives by modifying the model code of SASRec~\citep{kang2018self} and GRU4Rec~\citep{hidasi2015session}.
We choose SASRec and GRU4Rec for several reasons. 
\begin{enumerate*}[label=(\roman*)]
    \item SASRec and GRU4Rec are both state-of-the-art and widely-used encoders~\citep{wang2021personalized,klenitskiy2023turning}.
    \item RepeatNet is based on GRU4Rec.
    \item We want to compare the improvements over transformer-based encoders and RNN-based encoders.
\end{enumerate*}
\begin{itemize}[leftmargin=.2in,topsep=0pt]
\setlength\itemsep{0.0em}
\item \textbf{Softmax}: The performance of the SASRec and GRU4Rec. 
\item \textbf{Softmax + Mi}: Computing the probabilities using multiple input hidden states (Mi). Please see \Cref{sec:Mi} for more details. Here, Mi uses the hidden states corresponding to the last three input items and all the layers of the neural encoders (i.e., 1 layer for GRU4Rec and 2 layers for SASRec).
\item \textbf{Softmax + C}: Use context partition in \Cref{eq:logit_C}.
\item \textbf{Softmax + CP}: Use context partition and pointer network.
\item \textbf{Softmax + CPR:100}: Use softmax-CPR and set $k_1$ = 100 (i.e., removing the second and third reranker partition in \Cref{eq:logit_CPR}).
\item \textbf{Softmax + CPR:100 + Mi}: Use softmax-CPR and multiple input hidden states.
\item \textbf{Softmax + CPR:20,100,500 + Mi}: Use softmax-CPR in \Cref{eq:logit_CPR} and multiple input hidden states as in \Cref{fig:softmax_cpr}.
\item \textbf{Mixture of Softmax (MoS)}: The baseline similar to \citet{yang2018breaking,lin2021breaking}. We set the number of softmax to be 3.
\item \textbf{Softmax w/o Duplication~\citep{li2023repetition}}: Set the probability of the repeated items to be 0 via post-processing to improve the performance on the datasets without duplicated items.
\end{itemize} 
In addition, we also compare the softmax alternatives on top of SASRec/GRU4Rec with \textbf{RepeatNet}~\citep{ren2019repeatnet}. 

%In preliminary study on Steam, we find 20,100,500 is 

%We greatly accelerate the implementation of RepeatNet in RecBole, which also reduces its memory usage. 

%start our hyperparameter search from the default hyperparameters in RecBole and  . 

%using the validation set of every dataset .

%especially in smaller datasets. 
%hidden state

\subsection{Setup}

We report three metrics NDCG@10 (normalized discounted cumulative gain)~\citep{jarvelin2002cumulated}, HR@10 (Hit rate)~\citep{zhang2019deep}, and MRR@10 (Mean Reciprocal Rank)~\citep{radev2002evaluating}. To be closer to the real-world setup, we do not conduct negative example subsampling when reporting the testing performance, so the scores might look small in some datasets with a large number of items. Since different datasets could have very different performance ranges, we report the geometric mean of all datasets to summarize the performance of every method~\citep{valcarce2020assessing}.

We follow the default evaluation protocol and model setup in RecBole (e.g., input and output item embeddings are shared in GRU4Rec and SASRec). We found that the default hyperparameters in RecBole generally work well except that a smaller dropout rate for SASRec yields much better performances.
Overall, we found that our performance improvement is not sensitive to the hyperparameters but we still tried our best to tune the hyperparameters under the constraints of our computational resources.

For the smaller datasets (i.e., Amazon Beauty, Games, MovieLens 1m, and Steam), we perform a grid search on its hyperparameters using their NDCG@10 scores on validation sets. 
In the grid search, we use learning rates [5e-4, 1e-3, 2e-3] and batch sizes [64, 128]. For GRU4Rec, the dropout rates are [0, 0.5]. For SASRec, hidden state dropout rates are [0, 0.1]. For all the other 8 larger datasets, the learning rate is 1e-3, and dropout rate is 0. All the hyperparameter values or search ranges in RepeatNet are the same as GRU4Rec. To fitting the models into our GPU memory, we adjust our hidden state sizes and training batch sizes as shown in \Cref{tb:stats}. 

Using the grid search results, we can analyze the hyperparameter sensitivity of learning rates, dropouts, and batch sizes. To know the sensitivity to the hidden state sizes, we conduct another grid search using hidden state sizes [16, 32, 64, 128] and batch sizes [64,128]. The learning rate and dropout are set according to the previous grid search results to optimize the testing performance of the \textbf{Softmax + Mi} baseline in each of the 4 smaller datasets.

All experiments are done in Nvidia TESLA M40. The time is measured by computing the probability of all the items without using any nearest neighbor search. The model codes in RecBole are often not optimized for their running time, so the time comparison is more meaningful given the same neural encoder. Thus, we do not report the time of RepeatNet to avoid unfair comparisons.
%uses a very different model code, so its time is not comparable.
%compared to GRU4Rec

\begin{figure*}[t!]
\centering
\begin{subfigure}{.24\textwidth}
  \centering
  \includegraphics[width=1\linewidth]{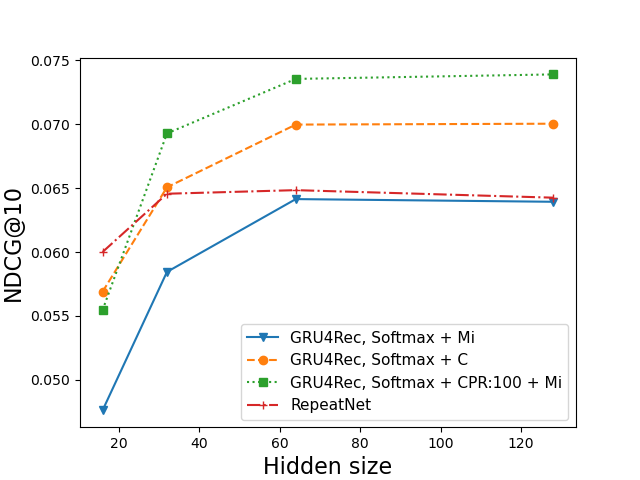}
  \caption{GRU encoder and $|\vh_{i_n}|$}
  \label{fig:gru_hidden}
\end{subfigure}%
\begin{subfigure}{.24\textwidth}
  \centering
  \includegraphics[width=1\linewidth]{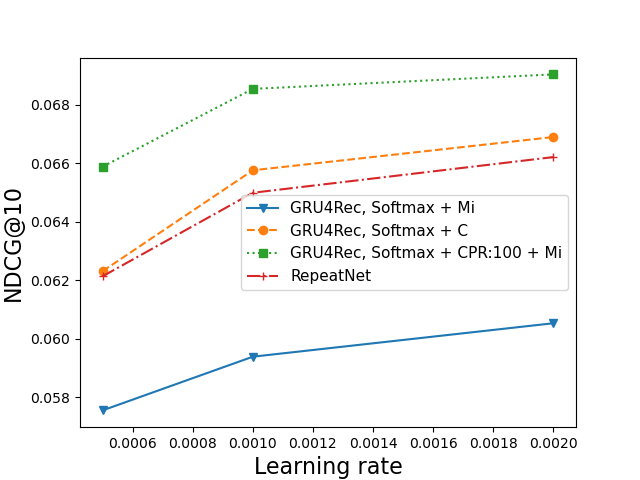}
  \caption{GRU encoder and lr}
  \label{fig:gru_lr}
\end{subfigure}
\begin{subfigure}{.24\textwidth}
  \centering
  \includegraphics[width=1\linewidth]{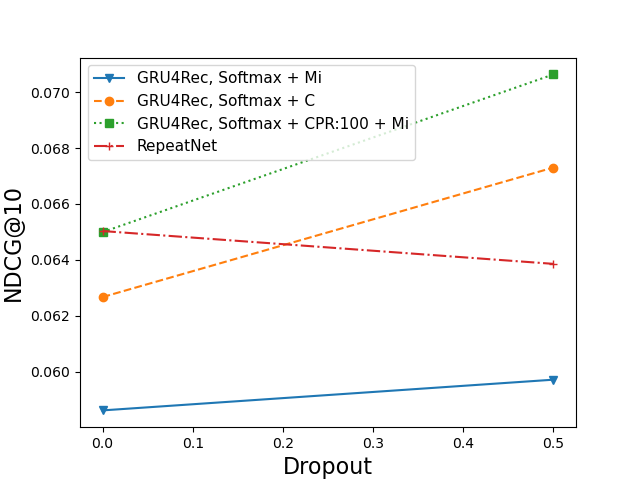}
  \caption{GRU encoder and dropout}
  \label{fig:gru_dropout}
\end{subfigure}
\begin{subfigure}{.24\textwidth}
  \centering
  \includegraphics[width=1\linewidth]{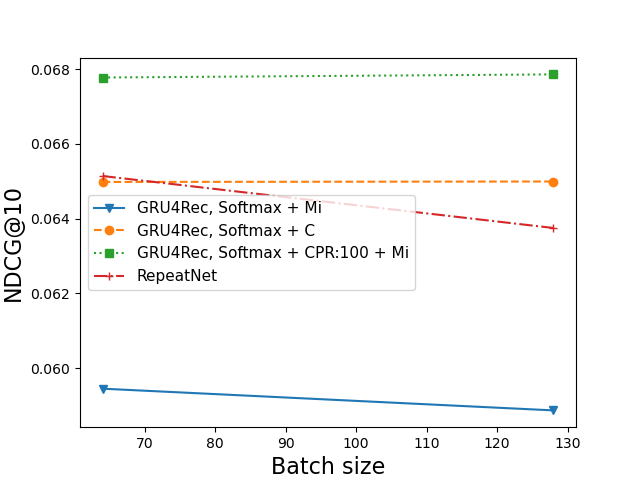}
  \caption{GRU encoder and batch size}
  \label{fig:gru_bsz}
\end{subfigure}
\begin{subfigure}{.24\textwidth}
  \centering
  \includegraphics[width=1\linewidth]{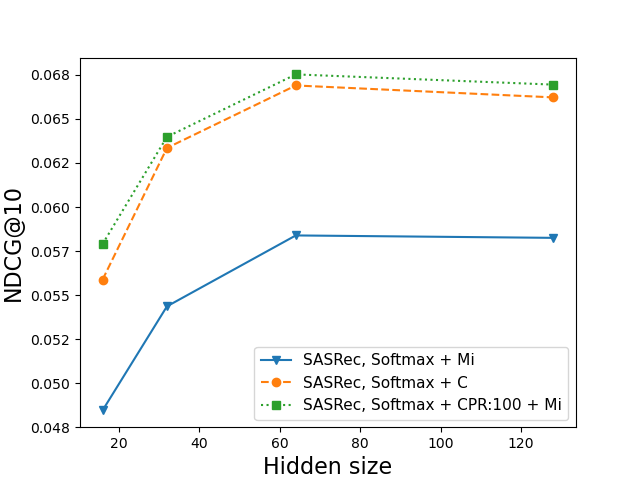}
  \caption{Trans. encoder and $|\vh_{i_n}|$}
  \label{fig:SAS_hidden}
\end{subfigure}
\begin{subfigure}{.24\textwidth}
  \centering
  \includegraphics[width=1\linewidth]{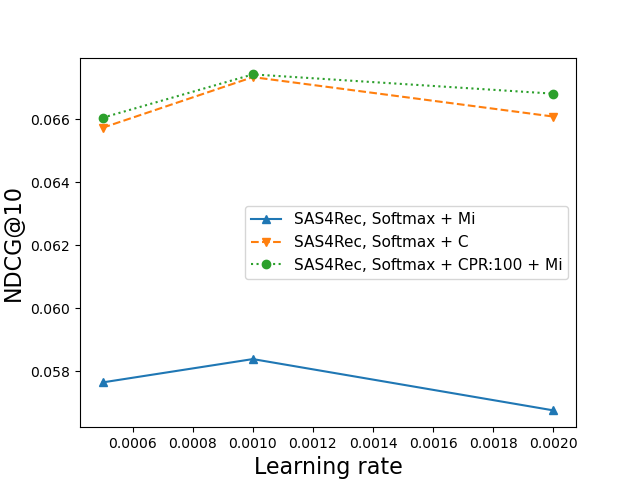}
  \caption{Trans. encoder and lr}
  \label{fig:SAS_lr}
\end{subfigure}
\begin{subfigure}{.24\textwidth}
  \centering
  \includegraphics[width=1\linewidth]{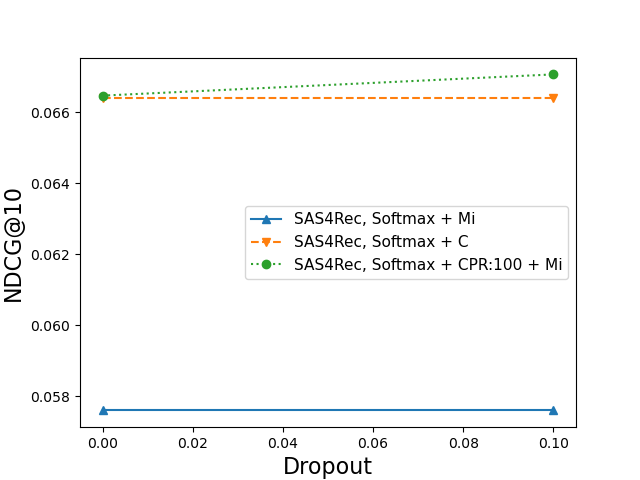}
  \caption{Trans. encoder and dropout}
  \label{fig:SAS_dropout}
\end{subfigure}
\begin{subfigure}{.24\textwidth}
  \centering
  \includegraphics[width=1\linewidth]{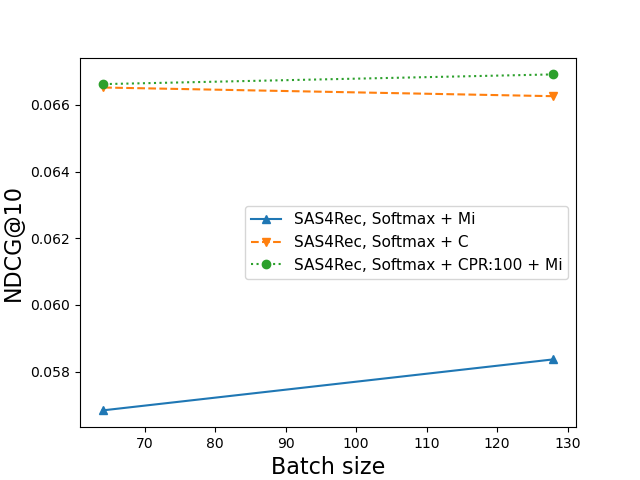}
  \caption{Trans. encoder and batch size}
  \label{fig:SAS_bsz}
\end{subfigure}
\caption{Hyerparameter analyses using the geometric mean of Amazon Beauty, Games, MovieLens 1m, and Steam datasets.}
\label{fig:hyper_analysis}
\end{figure*}

\subsection{Results}

The results are presented at \Cref{tb:nodup_dataset}, \Cref{tb:dup_dataset}, and \Cref{tb:all_datasets}. We can see context partition (\textbf{Softmax + C}) substantially improves over \textbf{Softmax}. After adding pointer network (P), reranker partition (R), and multiple input hidden states (Mi), \textbf{Softmax + CPR:100 + Mi} achieves the best overall performances in \Cref{tb:all_datasets}. Unlike the counterpart in language models~\citep{chang2023revisiting}, multiple reranker partitions, \textbf{Softmax + CPR:20,100,500 + Mi}, do not result in better performances in recommendation models.

The improvement on GRU4Rec is slightly larger than that on SASRec. This shows that there is a small overlap between the benefits of softmax-CPR and the benefits of self-attention in the neural encoder. Noting that the performances of SASRec and GRU4Rec are not directly comparable because we only coarsely tune the hyperparameters.
%the choose of neural encoder has minor effect on the choose of the softmax alternatives but 
%softmax-CPR reduces the benefits the self attention in the neural encoder.

\textbf{MoS}~\citep{yang2018breaking,lin2021breaking} performs almost the same as \textbf{Softmax} and requires much longer training and inference time. Although \textbf{Softmax w/o Duplication} performs poorly in the 5 datasets with duplications, it significantly outperforms \textbf{Softmax} in 7 datasets without duplications as found in \citet{li2023repetition}. Nevertheless, allowing the models to easily exclude the duplications during the training (e.g., \textbf{Softmax + C}) still slightly outperforms the post-processing baseline in datasets without duplications. 

We discover that although being designed for the datasets with duplicated items, \textbf{RepeatNet} can also substantially improve the datasets without any duplicated items. In many datasets, the performances of \textbf{RepeatNet} are very similar \textbf{Softmax + C} on top of GRU4Rec.
This suggests that the main source of improvement from \textbf{RepeatNet} comes from computing the probabilities of the repeated item separately as in \textbf{Softmax + C} rather than its self-attention mechanism or its extra parameters. Furthermore, after identifying the source of improvement from \textbf{RepeatNet}, \textbf{Softmax + C} can achieve similar improvement while only needing one-third of its model size, \textbf{Softmax + CPR:100 + Mi} can further expand the improvement by better overcoming the softmax bottleneck, and we can apply the softmax alternatives to any neural encoder of interest (e.g., they lead to similar improvement in SASRec). 

\Cref{tb:all_datasets} shows that the extra parameters introduced by our softmax alternatives are neglectable. Theoretically speaking, the extra computations in \textbf{Softmax + CPR:100 + Mi} are also very small compared to the original softmax layer, which computes the dot product between the hidden state and every item embedding. However, we still see some increases in training and testing time, which might be caused by the constraints of PyTorch's built-in functions. Thus, in applications requiring low latency, we recommend using \textbf{Softmax + C} and/or writing CUDA code to minimize the extra overhead.

The hyperparameter analyses in \Cref{fig:hyper_analysis} show that all the methods are not very sensitive to the particular values of hyperparameters. The lines are pretty flat in batch size and learning rate figures. For dropout rates in GRU4Rec, using $0$ dropout uniformly degrades the performance of smaller datasets such as Amazon Video Games. For hidden size, performance starts to degrade when the size is smaller than $64$. In RepeatNet, the item embedding size is two times of the hidden state size, so its $16$ hidden state size is similar to other methods' $32$ hidden state size.

%% file: content/related_work.tex
\section{Related Work}

Due to the commonness of repetition in sequential recommendation, many studies propose methods to improve the accuracy of recommending the repeated items. For example, \citet{bhagat2018buy,wang2019modeling} propose probabilistic models to find the proper time to recommend the items user bought before. \citet{ma2020modeling,ariannezhad2022recanet} propose special neural network architectures to explicitly model periodic user behavior. These methods usually design complicated models based on business insights in some specific domains, which might limit generalization ability and applicability to other domains.

\citet{li2023repetition,li2023next} systematically analyze the repetition and exploration behavior of existing sequential recommenders, and \citet{li2023repetition} observe that sharing the input and output embeddings could intensify the improper copying issue. Our work provides an explanation for the empirical observation (see \Cref{sec:bottleneck}).

\citet{yang2018breaking,chang2022softmax} introduce the concept of softmax bottleneck, the limitation of single embeddings, and the solution using multiple embeddings in MoS to improve language models. Recently, multiple embeddings are applied to information retrieval~\citep{khattab2020colbert,luan2021sparse,kong2022multi} and recommendation~\citep{wang2019attention,lin2021breaking}. Nevertheless, as we show in our experiment, the improvement of using multiple embeddings in MoS is often limited and inconsistent in sequential recommendation tasks.

Recently, \citet{chang2023revisiting} propose softmax-CPR to improve the distribution of the next word prediction and factuality of the generated text. Our work focuses on studying its meaning and effectiveness in the sequential recommendation tasks, which are previously unknown.

In our work, softmax-CPR uses different embeddings for the repetition intents and exploration intents. The idea is similar to multiple-intent recommendation or recommendation diversification. To diversify the recommendation, \citet{kula2017mixture} ensembles multiple LSTMs, \citet{kim2019sequential} clusters the items, and \citet{chen2021multi} formulates the sequential recommendation as a sequence to sequence task. Nevertheless, the diversity improvement often comes with much more complicated models specialized for specific datasets, limited recommendation accuracy improvement, and/or significantly increased computational overhead.

%% file: content/appendix.tex
\begin{figure*}[t!]
\centering
\begin{subfigure}{.45\textwidth}
  \centering
  \includegraphics[width=1\linewidth]{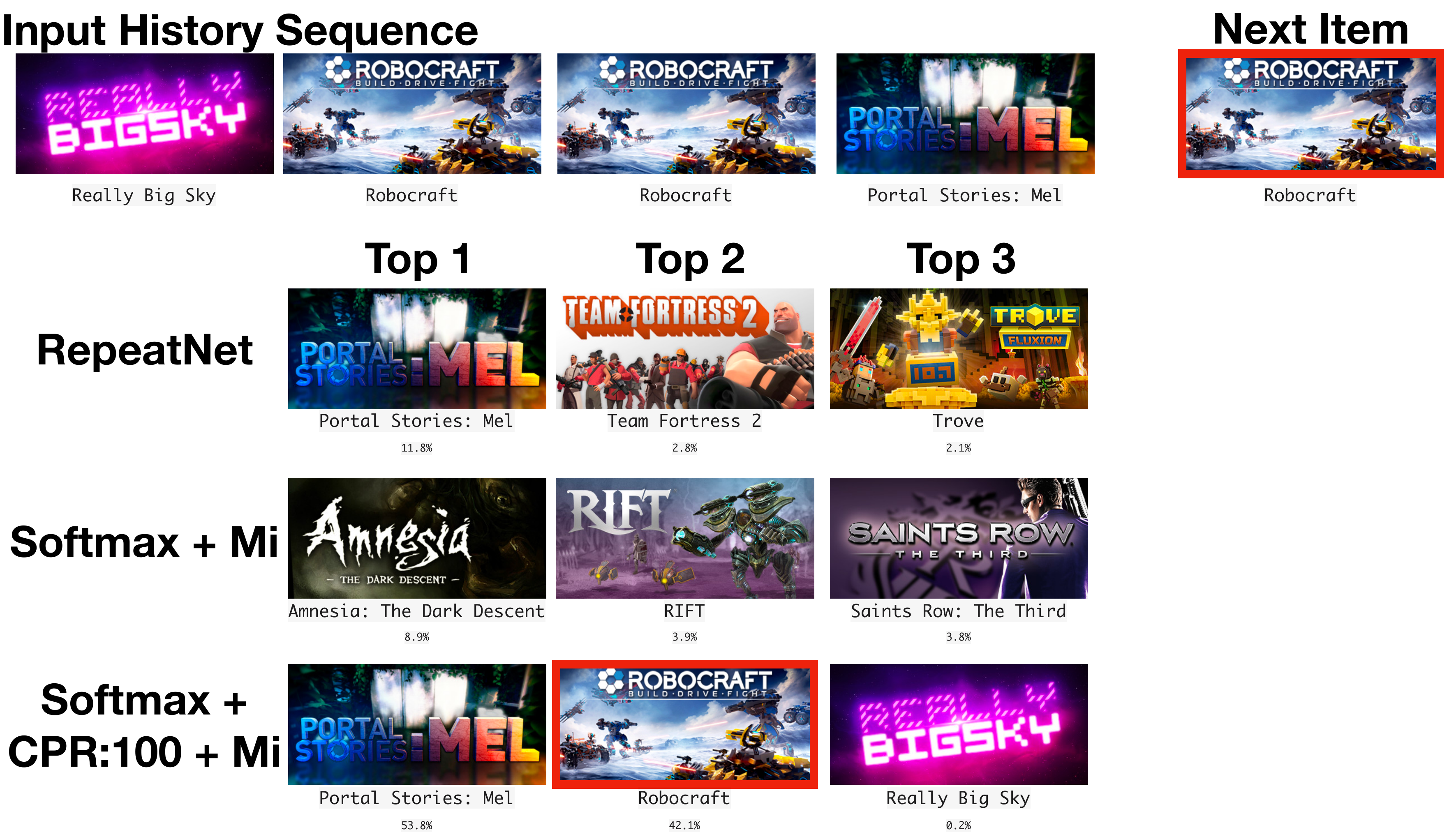}
  \caption{\textbf{Softmax + CPR:100 + Mi} learns to copy the input items correctly}
  \label{fig:repeat_example}
\end{subfigure} \hfill
\begin{subfigure}{.45\textwidth}
  \centering
  \includegraphics[width=1\linewidth]{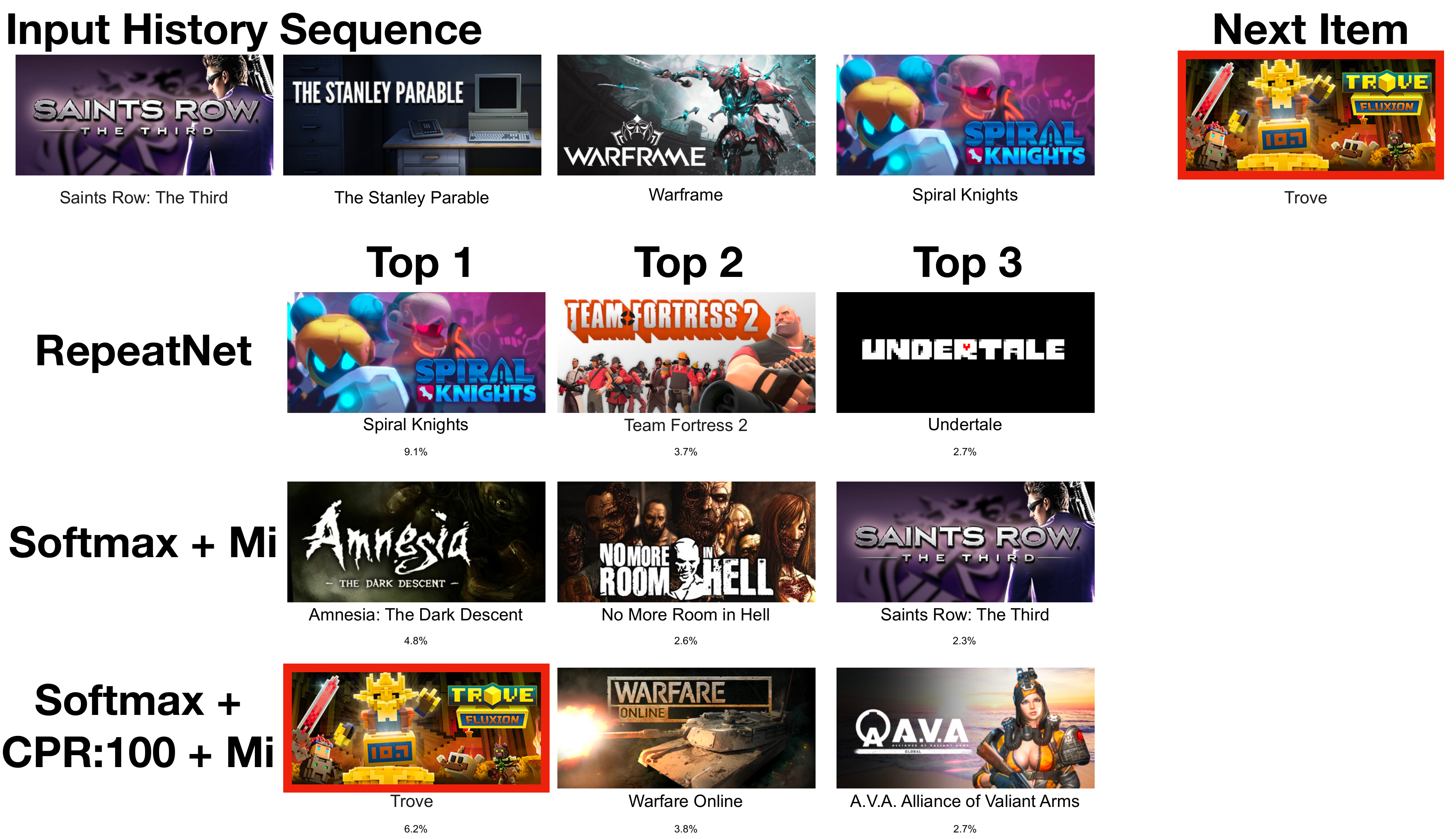}
  \caption{\textbf{Softmax + CPR:100 + Mi} learns to exclude the input items properly}
  \label{fig:not_repeat_example}
\end{subfigure}
\caption{Comparison of the Top 3 predictions in Steam dataset. The correct next item is highlighted using the red border.}
\label{fig:steam_examples}
\end{figure*}

\section{Overview}
In the supplementary material, we will describe our future work in \Cref{sec:future}, some details about our methods in \Cref{sec:method_details}, some details about our experiments in \Cref{sec:exp_details}, and present some examples in \Cref{sec:examples}. 

\section{Future Work}
\label{sec:future}
In this study, we studied the problem of using the single embedding in neural sequential recommenders that use cross-entropy loss. However, the problem should also exist for other losses such as BPR~\citep{rendle2009bpr} or other models based on matrix factorization \citep{rendle2020neural,ferrari2021troubling}. One important research question is whether the ideas of softmax-CPR could also significantly improve those methods.
% generally
%other recommendation models

The output softmax layer might also be a cause of lacking diversity in the recommended items because the item embeddings need to be close to the single embedding and thus close to each other. It is interesting to see if using softmax-CPR or other solutions could alleviate the problem.

Our current softmax-CPR implementation is written using pure Pytorch functions. Although the code is flexible and easy to maintain, they bring about some unnecessary computational overhead. We expect that writing the CUDA code directly should significantly speed up our softmax-CPR implementation.

For now, we only test our methods in the standard sequential recommendation settings using standard metrics. In the real world, even if recommending more repeated ones increases the prediction accuracy, it might not increase the long-term revenue because the users might lose chance of trying out new items. It would be interesting to see if the improvement gap changes in industrial recommendation systems (e.g., using larger dataset, nearest neighbor search, rerankers, and the revenue-oriented metrics). 

\section{Method Details}
\label{sec:method_details}
To make the results more comparable with the softmax-CPR, our \textbf{Softmax} baselines have one more linear projection layer for the hidden states $\vh_{i_n}$ compared to the standard GRU4Rec and SASRec. In our preliminary studies, this extra linear layer does not significantly change its performance.

\subsection{Softmax-CPR}
\label{sec:cpr_details}
In the main paper, we did not explain the pointer network and \textbf{Softmax + CP} in details. Here, we provide their formula. In \textbf{Softmax + CP}, we compute
\begin{equation}
\label{eq:logit_CP}
\text{Logit}_{CP}(x,i_n)=
\left\{
\begin{matrix*}[l]
\vf_{i_n,C}^T \vp_x + \vf_{i_n,P}^T \vf_{x,i_n,L} \;\; \text{if} \; x \in i_n\\[3pt] 
\vf_{i_n,V}^T \vp_x \;\; \text{O/W}
\end{matrix*}\right.,
\end{equation}
where $\vf_{i_n,P}=L_{P}^f(\vh_{i_n})$. In the softmax bottleneck section of the main paper, the main problem of the softmax layer is global item embedding $\vp_{.}$, so the pointer network directly solve this by predicting the local embeddings $\vf_{x,i_n,L}$ for the items in the input sequence. The local embedding of item $x$ are computed by
\begin{equation}
\label{eq:local_emb}
\vf_{x,i_n,L} = \frac{\sum_{j=1}^{n} \mathbbm{1}_{ {i_n^j}=x} L_{L}^f(\vq_{i_n^j}) }{\sum_{j=1}^{n} \mathbbm{1}_{{i_n^j}=x}},
\end{equation}
where $i_n^j$ is the $j$th input item in $i_n$, $L_{L}^f$ is the linear layer that projects the previous hidden states $\vq_{i_n^j}$ into the local item embeddings $\vf_{x,i_n,L}$, and $\mathbbm{1}_{ {i_n^j}=x} = 1 \; \; \text{if} \; i_n^j=x$. The performance on several pointer network variants in \citet{chang2023revisiting} show that the exact formulation here does not significantly affect the performance (e.g., whether we use summation or average).
%In the typical pointer networks such as RepeatNet,  
%We tried to use summation and average 

In the softmax-CPR formula in the main paper, 
%reranker partition
$P(k_2)$ is the top $k_2$ words with the highest logits, which come from $\vf_{i_n,R3}^T \vp_x$ if x is in $P(k_3)$ (the top $k_3$ words) and from $\vf_{i_n,V}^T \vp_x$ otherwise. $P(k_1)$ is the top $k_1$ words with the highest logits, which come from $\vf_{i_n,R2}^T \vp_x$ if x is in $P(k_2)$, from $\vf_{i_n,R3}^T \vp_x$ if x is in $P(k_3) - P(k_2)$, and from $\vf_{i_n,V}^T \vp_x$ otherwise.

%$P(k_1)$ is the top $k_1$ words with the highest $\max(\vf_{i_n,V}^T \vp_x, \vf_{i_n,R3}^T \vp_x)$

\subsection{Multiple Input Hidden States}

We expand the dimension of the hidden state by following the formula of \citet{chang2022softmax} 
\begin{equation}
\label{eq:multi-hidden}
\vq_{i_n} = \vh_{i_n}^M \oplus  GELU\left(L^h(\oplus_{j,m} \vh_{i_{n-j}}^{M-m})\right),
\end{equation}
where $GELU$ is a non-linear transformation~\citep{hendrycks2016gaussian}, $M$ is the number of layers in the neural encoder, $\oplus$ is concatenation, $\vh_{i_{n-j}}^{M-m}$ is the hidden state at $M-m$th layer corresponding to the $n-j$th input item, $L^h$ is a linear layer that reduces the size of the concatenated hidden state. For GRU4Rec, we consider the last 1x3 hidden states (i.e., hidden states for the last three items). For SASRec, we consider the last 2x3 hidden states.

%and $L^h$ is a linear transformation that allows us to consider more hidden states without significantly increasing the model size. $\oplus_{j,m} \vh_{i_{t-j}}^{M-m}$ is the concatenation of a block of hidden states. 
%We set the block size to be 3x3 in our GPT-2 experiments and 1x3 in our summarization experiments 
%(i.e., considering the last 3 hidden states in the last layer as shown in \autoref{fig:all_partition}).  

\subsection{Mixture of Softmax (MoS)}
In the standard MoS~\citep{yang2018breaking,lin2021breaking}, they usually predict the weights $\pi_{i_n,k}$ and perform the weighted average on the probability distribution from each softmax
\begin{equation}
\label{eq:MoS_org}
P^{Original}_{MoS}(x|i_n) = \sum\limits_{k=1}^K \pi_{i_n,k} \frac{\exp(\vf_{i_n,k}^T \vp_x)}{\sum_{x'} \exp(\vf_{i_n,k}^T \vp_{x'})}. 
\end{equation}

%use the following formula:
However, this approach makes nearest neighbor search very difficult. To accelerate the inference time, we just take the maximum on the logits from different hidden states.  %Then, one hidden state could be placed  
\begin{equation}
\label{eq:MoS}
P_{MoS}(x|i_n) =  \frac{\exp( \max_{k} \vf_{i_n,k}^T \vp_x)}{\sum_{x'} \exp( \max_{k} \vf_{i_n,k}^T \vp_{x'})}. 
\end{equation}
The original MoS is like a Gaussian mixture model and the efficient MoS is like a K-means. Both methods encourage each of their hidden states become the mode center of the ideal multi-modal distribution. In preliminary studies, their performances are also similar, so we use the efficient MoS method in our experiments.
%This is similar to original MoS because the highest logits would be emphasized after taking the exponential.
%\textit{apple juice} and \textit{diaper}

\subsection{RepeatNet}

The main reason why the number of parameters in RepeatNet would be 3 times of the GRU4Rec is that it does not share the input item embeddings and output item embeddings and the size of its output item embeddings is twice of the size of its input item embeddings and hidden states.

\section{Experiment Details}
\label{sec:exp_details}

For GRU4Rec, the number of layers is 1 and the item embedding size is 64. For SASRec, the number of layers is 2, number of heads is 2, inner size is 256, attention dropout rate is 0.1, and the item embedding size is the same as the hidden state size. The time measurement is done on the first 3 epochs of Amazon-2014 Books training and validation datasets. The hidden state size of all encoders is 64 when measuring the time. 

We set the number of reranking candidates as $k_1=20$, $k_2=100$, and $k_3=500$ using the validation set of Steam. In Bridge to Algebra (2008-2009) datasets, we use the problem step name as our item.

To understand the sensitivity of hyperparameters, we first fix the value of a target hyperparameter and compute the geometric means of NDCG@10 in the 4 smaller datasets in each configuration of the other hyperparameters. Then, we average the scores from all the other hyperparameters given each target hyperparameter value. When analyzing the sensitivity of hidden state size, GRU4Rec's item embedding size is the same as the hidden state size.

In the repetition statistics figures, we count the total number of the sequences with repetitions (blue) or without repetitions (orange) given a sequence length and sum over all the sequence lengths. We report the count in the legend of the figures.

\section{Real Examples and Analysis}
\label{sec:examples}
We visualize the top 3 recommendations from different methods in Steam dataset~\citep{kang2018self}. In the dataset, the items are video games and the interactions are the user review, representing the user's interest at that time. A user could leave multiple reviews for a game at different times. 

In \Cref{fig:repeat_example}, we can see that the user reviewed three games \textit{Really Big Sky}, \textit{Robocraft}, and \textit{Portal Stories: Mel} before and the user reviewed \textit{Robocraft} twice. \textbf{Softmax + CPR:100 + Mi} predicts that the user is very likely to review the last two games again and less likely to review the first game again, which makes sense because \textit{Really Big Sky} is not a well-known game. \textbf{RepeatNet} only predicts that user might repeat the last game and \textbf{Softmax + Mi} does not predict any repeated item in its top 3 recommendation list. 

In \Cref{fig:not_repeat_example}, the user has a broad interest and did not reviewed the same game twice. \textbf{RepeatNet} still predicts that user might repeat the last game and \textbf{Softmax + Mi} copies \textit{Saints Row: The Third}, the first game user reviewed. The other two recommendations from \textbf{Softmax + Mi} are similar to \textit{Saints Row: The Third}. The top three predictions from \textbf{Softmax + CPR:100 + Mi} are three popular free-to-play games, \textit{Trove}, \textit{Warfare Online}, and \textit{A.V.A. Alliance of Valiant Arms}. These recommendations are reasonable because the last two games the user reviewed, \textit{Warframe} and \textit{Spiral Knights}, are also popular free-to-play games.